\documentclass[prx,twocolumn,showpacs,amsmath,amssymb,superscriptaddress]{revtex4-2}

\usepackage{amsmath,amssymb,amsfonts,bm,color,graphicx,tabularx}

\usepackage[unicode=true,colorlinks=true]{hyperref}

\usepackage[etex=true,export]{adjustbox}
\usepackage{makecell}
\usepackage{multirow}
\usepackage{amsmath}
\usepackage{comment}

\hypersetup{linkcolor=blue, citecolor=blue,urlcolor=blue}
\usepackage{tabularx,multirow,array,diagbox}
\usepackage{adjustbox}

\renewcommand{\v}[1]{{\bf #1}}

\newcommand{\beq}{\begin{equation}}
\newcommand{\eeq}{\end{equation}}
\newcommand{\beql}{\begin{equation*}}
\newcommand{\eeql}{\end{equation*}}
\newcommand{\beqn}{\begin{eqnarray}}
\newcommand{\eeqn}{\end{eqnarray}}

\begin{document}
\title{Boundary topological insulators and superconductors of Altland-Zirnbauer tenfold  classes}
\author{Xun-Jiang Luo}
\affiliation{Department of Physics, Hong Kong University of Science and Technology, Clear Water Bay, Hong Kong, China}
\affiliation{School of Physics and Technology, Wuhan University, Wuhan 430072, China}
\author{Fengcheng Wu}
\email{wufcheng@whu.edu.cn}
\affiliation{School of Physics and Technology, Wuhan University, Wuhan 430072, China}
\affiliation{Wuhan Institute of Quantum Technology, Wuhan 430206, China}

\begin{abstract}

%\date{\today}

In a class of systems, there are gapped boundary-localized states described by a boundary Hamiltonian.
 The topological classification of gapped boundary Hamiltonians, same as the standard tenfold way for gapped bulk states,  can lead to the emergence of boundary topological insulators (TIs) and superconductors (TSCs). In this work, we present a theoretical study of boundary TIs and TSCs of the full Altland-Zirnbauer tenfold symmetry classes. Based on the boundary projection analyses for a $d$-dimensional Dirac continuum model, we demonstrate that nontrivial boundary topology can arise at a $(d-n)$-dimensional boundary if the Dirac model incorporates ($n+1$) mass terms with $0<n<d$ although its bulk and $(d-1)$D,$\cdots$, $(d-n+1)$D boundaries are topologically trivial. Furthermore, we present a unified criterion for the emergence of nontrivial boundary topology by extending bulk classification within the context of the Dirac model, which provides a unified framework for nontrivial bulk and boundary topology.  Inspired by the Dirac continuum model analysis, we further construct bulk lattice Hamiltonians for realizing boundary  TIs and TSCs of the full Altland-Zirnbauer tenfold symmetry classes,  which enables the realization of higher-order TIs and TSCs in arbitrary dimensions with arbitrary orders.  We analyze some typical examples of the constructed boundary TIs and TSCs in physical dimensions.

\end{abstract}
\maketitle

\section{Introduction}
\begin{figure}
\centering
\includegraphics[width=3.3in]{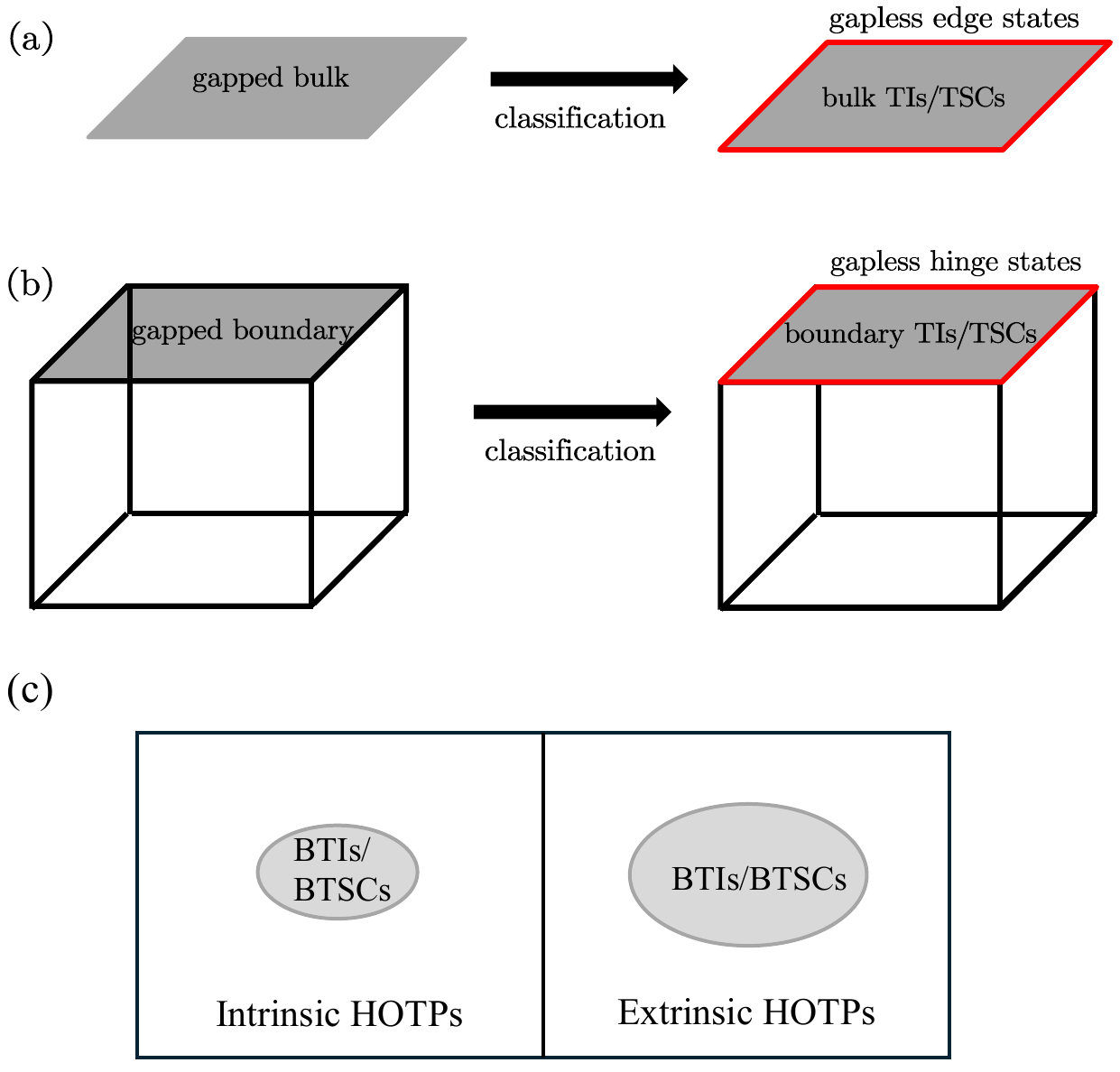}
\caption{(a) The schematic illustration of obtaining bulk TIs/TSCs by performing topological classification of all the gapped bulk systems. The bulk
TIs/TSCs feature gapless boundary states at their $(d-1)$D boundaries. (b) The schematic illustration of obtaining boundary TIs/TSCs by performing topological classification of the gapped boundary. The boundary TIs/TSCs feature gapless boundary states of codimension larger than one. (c) The Venn diagram for boundary TIs /TSCs (BTIs/BTSCs) and HOTPs.}
\label{fig1}
\end{figure}

Since the discovery of time-reversal invariant TIs \cite{Kane2005,Kane2005a,Bernevig2006,Koenig2007}, topological phases of matter have emerged as one of the most vibrant fields in physics \cite{Hasan2010,Qi2011}. TIs and TSCs are characterized by a gapped bulk energy spectrum accompanied by gapless boundary states. These gapless boundary states can be protected by the presence of local symmetries, including time-reversal symmetry
$\mathcal{T}$, particle-hole symmetry $\mathcal{P}$, and/or their combination chiral symmetry $\mathcal{C}$. According to the presence
or absence of these symmetries, all the gapped systems can be
classified into the Altland-Zirnbauer (AZ) tenfold symmetry classes \cite{Altland1997}. The  classification of gapped Hamiltonians leads to the well-known topological periodic
tables of
TIs and TSCs \cite{Schnyder2008,Kitaev2009,Ryu2010,Chiu2016}.   Interestingly, when the classified gapped Hamiltonians describe some systems's boundaries, this standard tenfold classification can naturally lead to boundary TIs and TSCs characterized by the gapless boundary states at the boundary of the boundary, as schematically illustrated in Figs.~\ref{fig1}(a)-(b).

The quest for robust gapless boundary states of codimension greater than one, such as corner states and hinge states, has spurred the discovery of higher-order topological phases (HOTPs) \cite{Volovik2010,Sitte2012,Zhang2013a,Benalcazar2017a,Benalcazar2017,Song2017,Langbehn2017,Schindler2018,Geier2018,Khalaf2018,Khalaf2018a}. In recent years, extensive research has been devoted to higher-order TIs \cite{Song2017,Langbehn2017,Khalaf2018a,Schindler2018,Geier2018,Khalaf2018,Ezawa2018a,Ezawa2018,Schindler2018a,Xu2019,Yue2019,Sheng2019,Park2019,Chen2020,Ren2020,Zhang2020b,Liu2021,Jia2023,Jiazheng2024,2025arXivL} and TSCs \cite{Yan2018,Wang2018,Zhu2018,Hsu2018,Pan2019,Yan2019,Zhang2019,Zhu2019,Zhang2019a,Volpez2019,Ahn2020,Wu2020,Kheirkhah2020,Wu2020a,Hsu2020,Pan2021,chenli2021,Zhang2021,Tan2022,ZhangZhongyi2022,PhysRevB.111.184516,Zhang2024,zhang2024topological,zhang2024fermi} across various systems,  including electronic \cite{Schindler2018a,Yue2019,Xu2019,Sheng2019,Park2019,Chen2020,Ren2020,Zhang2020b,Liu2021}, bosonic \cite{Xie2018,Serra-Garcia2018,Xie2019,Chen2019c,Ni2019,Fan2019,Xue2019}, Floquet \cite{Rodriguez-Vega2019,Nag2019,Peng2019a,Peng2020,Hu2020,Huang2020,Nag2021,Ghosh2021a}, non-Hermitian \cite{Liu2019,Zhang2019d,Luo2019,Edvardsson2019,Kawabata2020}, and quasicrystal \cite{Varjas2019,Chen2020a,Hua2020,Peng2021,Spurrier2020,Lv2021,Wang2022} systems. An $n$th-order topological phase in a $d$-dimensional ($d$D) system is characterized by the presence of robust gapless boundary states at its ($d-n$)D boundaries. In this context, conventional TIs and TSCs are classified as first-order phases. Generally, HOTPs can be categorized into two distinct types \cite{Geier2018,Trifunovic2019}: (i) intrinsic HOTPs, which are associated with crystalline-symmetry-protected bulk topology \cite{Song2017,Langbehn2017,Shiozaki2017,Khalaf2018a,Schindler2018,Geier2018,Khalaf2018,Huang2021}; and (ii) extrinsic HOTPs, whose bulk is topologically trivial but features topological boundary states with a codimension large than one 
\cite{Khalaf2021}.
 Notably, for a 
$d$D $n$th-order topological phase, whether intrinsic or extrinsic, its 
($d-n+1$)D boundaries generally  play a crucial role in generating the
($d-n$)D gapless boundary states.
There is a special type of HOTPs, named boundary TIs or TSCs, where  a boundary Hamiltonian fully describes gapped states localized at the ($d-n+1$)D boundary and its topology gives rise to the $(d-n)$D gapless boundary states.

A prominent example of HOTPs is the Benalcazar-Bernevig-Hughes (BBH) model, introduced in the seminal works of Ref. \onlinecite{Benalcazar2017a,Benalcazar2017}, which initiated extensive exploration of HOTPs. Recently, a study \cite{Luo2024} generalized the 2D BBH model to arbitrary dimensions and demonstrated that the 1D boundaries of the generalized BBH model can be exactly described by the two-band Su-Schrieffer-Heeger (SSH) model, which serves as a minimal model for realizing first-order TIs. Consequently, both the BBH model and its higher-dimensional generalizations can be regarded as paradigms of boundary TIs within the AIII symmetry class with chiral symmetry. Notably, this boundary TI is well-defined because the 1D boundary states of the generalized BBH model extend throughout the entire boundary Brillouin zone, and therefore allow for a lattice model description for boundary states \cite{Luo2024,LuoXunJiang2025}. This situation sharply contrasts with the gapped boundary states observed in the majority of HOTPs \cite{Chen2020,Ren2020,Zhang2020b,Yan2018,Wang2018,Zhu2018}, which are typically inherited from first-order TIs and TSCs and consequently lack a lattice model description. On the other hand, several studies \cite{Chen2019a,Li2021a,Luo2021a} proposed to realize boundary TIs and TSCs in solid systems, including surface Chern insulators \cite{Chen2019a} and surface TSCs \cite{Li2021a,Luo2021a}.

In this work, we present a systematic study of boundary topological TIs and TSCs of the full AZ tenfold symmetry classes. By performing boundary projection analysis for the $d$D Dirac continuum model with $(n+1)$ mass terms ($0<n<d$), we obtain a ($d-n$)D boundary Dirac Hamiltonian with single Dirac mass term, which is known to describe nontrivial topology in the absence of a symmetry-preserving extra mass term (SPEMT) \cite{Morimoto2013,Chiu2013,Chiu2014,Chiu2014a}. We find that this condition can be satisfied if the bulk Hamiltonian lacks SPEMT, thereby presenting  a unified criterion
for the emergence of nontrivial boundary topology. This criterion  provides a unified framework for non-trivial bulk and boundary topology within the Dirac model.
Inspired by the continuum model analysis, we further construct boundary TIs and TSCs of the full AZ symmetry classes, which are described by a boundary Dirac lattice Hamiltonian
defined on the whole boundary Brillouin zone. Our constructions naturally lead to the HOTPs in arbitrary dimensions with arbitrary orders and we derive the analytical solution of the gapless boundary states. Finally, 
some typical examples
of the constructed boundary TIs and TSCs in physical dimensions are discussed. Our work advances the study of boundary topological phases by offering a unified framework for both physical understanding and model construction.

We note that when crystal symmetry is not considered, the gapless boundary states of boundary TIs or TSCs can be removed by closing the boundary energy gap. In this case, boundary TIs and TSCs fall into the classification of extrinsic HOTPs. However, when certain crystal symmetries are considered, boundary TIs can also be intrinsic HOTPs. An example is the two-dimensional Benalcazar-Bernevig-Hughes model with four-fold rotation symmetry, as discussed in Sec.~\ref{VI}. In Fig.~\ref{fig1}(c), we schematically illustrate the relations between boundary TIs/(TSCs) and higher-order TIs/(TSCs).

This paper is organized as follows. In Sec.~\ref{II},  we briefly
review the application of the Dirac model in the classification
 and description of first-order TIs and TSCs.  In Sec.~\ref{III}, we perform boundary projection analyses for the Dirac continuum model with multi-mass terms to demonstrate the existence of nontrivial boundary topology and present a unified criterion for their emergence. In Sec.~\ref{IV}, we construct lattice models of boundary TIs and TSCs of the full AZ symmetry classes. In Sec.~\ref{V}, we present some typical examples of the constructed boundary TIs and TSCs in physical dimensions. In Sec.~\ref{VI}, we give a brief discussion and summary. Appendices \ref{Appendix A}-\ref{Appendix D} complement the main text.

\section{Review of Dirac Hamiltonian}
\label{II}

In order to introduce concepts that are essential to this work, we first review the topological classification of first-order TIs and TSCs using the Dirac Hamiltonian.
For a $d$D system, the Dirac Hamiltonian with a single Dirac mass can be written as
\beqn
H_d^{(1)}(\bm k_d)=\sum_{i=1}^d k_i \gamma_i^{(g)}+m_0 \gamma_{d+1}^{(g)},
\label{dh}
\eeqn
which is supposed to describe the gap-closing and reopening process of any gapped systems \cite{Morimoto2013,Chiu2013,Chiu2014,Chiu2014a}.  The momentum vector is $\bm k_d=(k_1,\cdots,k_d)$ with $k_i$ being the momentum component of the ${i}$-th direction and $m_0$ denotes the mass.
 The superscript and subscript of  $H_d^{(1)}$ denote the number of mass terms and the dimension of the system, respectively. We use this convention throughout this paper. The
$2^g\times 2^g$ Gamma matrices $\gamma_j^{(g)}$ (see Appendix \ref{Appendix A}) obey the anti-commutation relation $\left\{\gamma_j^{(g)}, \gamma_{j^{\prime}}^{(g)}\right\}=2 \delta_{jj^{\prime}}I^{(g)}$ with $j,j^{\prime}=1,\cdots,d+1$, where $I^{(g)}$ is $2^g\times 2^g$ identity matrix. Since there is a maximum of $(2g + 1)$ anti-commuting gamma matrices with the dimension $2^g\times 2^g$, we require that $d \leq 2g$ in $H_d^{(1)}$.
The three local symmetries $\mathcal{T}$, $\mathcal{P}$, and $\mathcal{C}$, if present, act on $H_d^{(1)}$ as
\beqn
&&\mathcal{T}H_d^{(1)}(\bm k_d)\mathcal{T}^{-1}=H_d^{(1)}(-\bm k_d),\nonumber\\
&&\mathcal{P}H_d^{(1)}(\bm k_d)\mathcal{P}^{-1}=-H_d^{(1)}(-\bm k_d),\nonumber\\
&&\mathcal{C}H_d^{(1)}(\bm k_d)\mathcal{C}^{-1}=-H_d^{(1)}(\bm k_d),
\eeqn
which lead to
\beqn
[\gamma_{d+1}^{(g)},\mathcal{T}]=0,\{\gamma_{1,\cdots,d}^{(g)},\mathcal{T}\}=0,\nonumber\\
\{\gamma_{d+1}^{(g)},\mathcal{P}\}=0,[\gamma_{1,\cdots,d}^{(g)},\mathcal{P}]=0,\nonumber\\
\{\gamma_{d+1}^{(g)},\mathcal{C}\}=0,\{\gamma_{1,\cdots,d}^{(g)},\mathcal{C}\}=0.
\label{cr}
\eeqn
The presence or absence of the three local symmetries classifies all the gapped systems into the AZ tenfold symmetry classes.

The topological classification of gapped systems of each symmetry class is completely determined by the existence of SPEMT for $H_d^{(1)}$ \cite{Morimoto2013,Chiu2013,Chiu2014,Chiu2014a}, labeled as $m_1\gamma_{d+2}^{(g)}$, where $\{\gamma_{d+2}^{(g)}, H_d^{(1)}\}=0$ and $\gamma_{d+2}^{(g)}$ satisfies the same commutation
 relations given by Eq.~\eqref{cr} as $\gamma_{d+1}^{(g)}$. If
SPEMT exists, the phases of $H_d^{(1)}$ characterized by $m_0>0$ and $m_0<0$ can be continuously connected without breaking any symmetries and closing the bulk energy gap. In this scenario, $H_d^{(1)}$ describes only one phase, namely atomic insulator.  Conversely, if SPEMT does not exist, phases characterized by $m_0>0$, and $m_0<0$ are topologically distinct, and $H_d^{(1)}$ has a $Z$ or $Z_2$ topological character, which can be further determined by examining the existence of SPEMT within a larger system that comprises two copies of $H_d^{(1)}$. 
If SPEMT exists for this doubled system, $H_d^{(1)}$ has a $Z_2$ topological classification. If not, $H_d^{(1)}$ has a $Z$ topological classification. 
This classification scheme ultimately produces the classification periodic table of first-order TIs and TSCs (see Appendix~\ref{appendix e}).

Dirac model also provides a unified description of first-order TIs and TSCs. 
By replacing $k_i$ with $\sin k_i$ and $m_0$ with $M_0(\bm k_d)$ in Eq.~\eqref{dh}, we obtain the Dirac lattice Hamiltonian
\beqn
{H}_{d}^{(1)}(\bm k_d)=\sum\limits_{i=1}^{d}\sin k_i\gamma_{i}^{(g)}+M_0(\bm k_d)\gamma_{d+1}^{(g)},
\label{Dirac}
\eeqn
where $M_0(\bm k_d)=\mathcal{M}+\sum_{i=1}^{d}(1-\cos k_i)$ with $-2<\mathcal{M}<0$ ensuring a band inversion at the momentum zero point  $\bm k_d=0$. For a given symmetry class and dimension $d$, if ${H}_{d}^{(1)}$ lacks SPEMT, ${H}_{d}^{(1)}$ describes a first-order TI or TSC. This is because the gapped phase in such a configuration cannot be continuously deformed into the topologically trivial atomic limit ($M_0(\bm k_d) = \mathcal{M}$) without closing bulk energy gap or breaking any symmetries. Here we list several representative examples of TIs described by ${H}_{d}^{(1)}$. When $d=1$ and $g=1$, ${H}_{d}^{(1)}$ is the two-band SSH model acting as a 1D TI in the AIII symmetry class; when $d=2$ and $g=1$, ${H}_{d}^{(1)}$ denotes the Qi-Wu-Zhang model \cite{Qi2006} and describes a Chern insulator in the A symmetry class; when $d=2$ and $g=2$, ${H}_{d}^{(1)}$ is the Bernevig–Hughes–Zhang model \cite{Bernevig2006} and describes a time-reversal invariant TI in the AII symmetry class; when $d=3$ and $g=2$, ${H}_{d}^{(1)}$ describes a 3D TI in the AII symmetry class.

\begin{table*}[htb]
\centering
\setlength\tabcolsep{4.7pt}
\renewcommand{\multirowsetup}{\centering}
\renewcommand{\arraystretch}{1.6}
\caption{Minimal matrix dimension of ${H}_{d}^{(n+1)}$ that realizes boundary TIs and TSCs. $\tilde{H}_q^{(1)}$ is the obtained $(d-n)$D boundary Hamiltonian from ${H}_{d}^{(n+1)}$ with $d=n+q$ and describes the first-order TIs or TSCs in spatial dimension $q$. The minimal matrix dimension of $\tilde{H}_q^{(1)}$ in different symmetry classes is obtained from Ref.~\onlinecite{Chiu2014a}.}
\begin{tabular}{|c|c|c|c|c|c|c|c|c|c|c|c|c|c|c|c|}
\hline
Spatial dimension of $\tilde{H}_{q}^{(1)}$ &\multicolumn{5}{|c|}{$q=1$}&\multicolumn{5}{|c|}{$q=2$}&\multicolumn{5}{|c|}{$q=3$}\\
\hline
\diagbox{matrix dimension}{symmetry class}&AIII&D & DIII & CII & BDI&A&D&DIII&AII&C&AIII&DIII&AII&CII&CI\\
\hline
size($\tilde{H}_{q}^{(1)}$)&$2^{1}$&$2^{1}$ & $2^{2}$ &$2^{2}$ &$2^{1}$&$2^{1}$&$2^{1}$&$2^{2}$&$2^{2}$&$2^{2}$&$2^{2}$&$2^{2}$&$2^{2}$&$2^{3}$&$2^{3}$\\
\hline
size($H_{d}^{(n+1)}$)&\multicolumn{15}{|c|}{$2^{n}\times \text{size}(\tilde{H}_q^{(1)})$}\\
\hline
\end{tabular}
\label{tabtf}
\end{table*}

\section{Nontrivial Boundary topology}
\label{III}
Since $H_d^{(1)}$ can be topologically nontrivial, nontrivial boundary topology can emerge when the boundary of a system is described by this Hamiltonian. In the following,  we demonstrate that nontrivial boundary topology can arise by introducing additional mass terms based on $H_d^{(1)}$. To effectively capture the essence of topology, we primarily focus on the continuum model Hamiltonian in this section.

\subsection{Nontrivial topology on $(d-1)$D boundaries}
\label{IIIA}
We add an additional mass term $m_1\gamma_{d+2}^{(g)}$ to $H_d^{(1)}$ and then the obtained new Hamiltonian can be written as
\beqn
{H}_d^{(2)}(\bm k_d)=\sum_{i=1}^d k_i \gamma_i^{(g)}+m_0 \gamma_{d+1}^{(g)}+m_1\gamma_{d+2}^{(g)}.
\label{2m}
\eeqn
Since there are two mass terms for ${H}_d^{(2)}$, all the gapped phases described by ${H}_d^{(2)}$ can always be continuously deformed into each other without closing the bulk energy gap. Therefore, ${H}_d^{(2)}$ describes the same phase as the atomic insulator. However, we claim that the $(d-1)$D boundaries of this system can possess nontrivial topology if ${H}_d^{(2)}$ does not allow SPEMT, although the bulk is topologically trivial.

To substantiate this claim, we perform surface projection analysis by considering a mass domain wall characterized by $m_0(r_d)$ along the $r_d$ direction. $m_0(r_d)$ is defined as  
$m_0(r_d)=m_0>0$ in the region $r_d > 0$ and $m_0(r_d)=-m_0<0$ in the region $r_d <0$. In this case, $k_d$ is not a good quantum number and we replace $k_d$ by $-i\partial_{r_d}$. We separate $H_d^{(2)}$ into two parts
\beqn
&&H_d^{(2)}={H}_{d-1}^{(1)}(\bm k_{d-1})+{H}_1^{(1)}(r_d),\nonumber\\
&&{H}_{d-1}^{(1)}(\bm k_{d-1})=\sum_{i=1}^{d-1}k_i\gamma_i^{(g)}+m_1\gamma_{d+2}^{(g)},\nonumber\\
&&{H}_1^{(1)}(r_d)= -i\gamma_d^{(g)}\partial_{r_d}+m_0(r_d)\gamma_{d+1}^{(g)},
\label{hd0}
\eeqn
where $\bm k_{d-1}=(k_1,\cdots,k_{d-1})$.
It can be shown that ${H}_1^{(1)}(r_d)$ hosts the zero-energy states (see Appendix \ref{Appendix B}),
\beqn
\Psi_{-}(r_d)=\mathcal{N}e^{-\int_{0}^{r_d}m_0(r_d^{\prime})dr_d^{\prime}}\psi_{-},
\label{dw1}
\eeqn
where spinor $\psi_{-}$ is the eigenstate of $i\gamma_d^{(g)}\gamma_{d+1}^{(g)}$ with eigenvalue $-1$, namely $i\gamma_d^{(g)}\gamma_{d+1}^{(g)}\psi_{-}=-\psi_{-}$,  and $\mathcal{N}$ is the normalization factor. By projecting ${H}_{d-1}^{(1)}$ onto the subspace spanned by  $\Psi_{-}(r_d)$, we obtain the boundary Hamiltonian
\beqn
\tilde{{H}}_{d-1}^{(1)}=P_-{H}_{d-1}^{(1)}P_-=\sum_{i=1}^{d-1}k_i\tilde{\gamma}_i^{(g-1)}+m_1\tilde{\gamma}_{d+2}^{(g-1)},
\label{bh}
\eeqn
where the projection operator is defined as $P_-=(1-i\gamma_d^{(g)}\gamma_{d+1}^{(g)})/2$. $\tilde{\gamma}_j^{(g-1)}$ has a half matrix size compared to that of
${\gamma}_j^{(g)}$ when picking the nonzero block part and satisfies the anti-commutation relation
\beqn
\left\{\tilde{\gamma}_j^{(g-1)}, \tilde{\gamma}_{j^{\prime}}^{(g-1)}\right\}=2 \delta_{ jj^{\prime}}I^{(g-1)},
\eeqn
with $j,j^{\prime}=1,\cdots,d-1,d+2$.
It is noteworthy that the boundary Hamiltonian $\tilde{{H}}_{d-1}^{(1)}$ maintains a similar form as $H_d^{(1)}$ (Eq.~\eqref{dh}) with a single mass term. Therefore,
the topological classification scheme presented for $H_d^{(1)}$ is also applicable to 
$\tilde{{H}}_{d-1}^{(1)}$, which implies that $\tilde{{H}}_{d-1}^{(1)}$ can exhibit nontrivial topology if it lacks SPEMT.
However, examining this condition is complicated since obtaining $\tilde{{H}}_{d-1}^{(1)}$ involves the boundary projection process. In the following, we present the criterion for the emergence of nontrivial boundary topology for the bulk Hamiltonian ${H}_{d}^{(2)}$.

Since local symmetries do not act on spatial coordinates,  the boundary Hamiltonian $\tilde{H}_{d-1}^{(1)}$ belongs to the same symmetry class as the bulk Hamiltonian $H_d^{(2)}$,
which can be
explicitly shown by considering the concrete representation
 of the Gamma matrices (see Appendix \ref{Appendix C}).
 In Appendix \ref{Appendix C} , we establish a one-to-one correspondence between the SPEMT $m_2\tilde{\gamma}_{d+3}^{(g-1)}$ for $\tilde{{H}}_{d-1}^{(1)}$ and the SPEMT $m_2\gamma_{d+3}^{(g)}$ for $H_d^{(2)}$. 
This correspondence implies that if ${H}_d^{(2)}$ lacks a SPEMT, then $\tilde{{H}}_{d-1}^{(1)}$ must also lack one. In this scenario, the boundary phases described by $\tilde{{H}}_{d-1}^{(1)}$ with $m_1>0$ and $m_1<0$ are topologically distinguished, and $\tilde{{H}}_{d-1}^{(1)}$ has a $Z$ or $Z_2$ topological character, which can be further determined by examining the existence of SPEMT for two copies of $\tilde{{H}}_{d-1}^{(1)}$ or ${H}_d^{(2)}$.
Thus, we can make a remarkable conclusion that
when ${H}_d^{(2)}$ does not allow SPEMT, although the bulk is topologically trivial, its $(d-1)$D boundaries can possess nontrivial topology. 
In the following, we present two examples involving the second-order TIs phase to illustrate the established criterion for the emergence of nontrivial boundary topology.
  
The first example is a second-order TI featuring hinge states, which are realized by introducing an addition mass term to gap the surface states of a 3D TI \cite{Schindler2018}. The low-energy effective Hamiltonian for this model can be expressed as 
\beqn
H_3^{(2)}=k_1\gamma_1^{(2)}+k_2\gamma_2^{(2)}+k_3\gamma_3^{(2)}+m_0\gamma_4^{(2)}+m_1\gamma_{5}^{(2)},
\eeqn
which belongs to the A symmetry class. It can be verified that both
$H_3^{(2)}$ and its double do not allow SPEMT, which indicates that the surfaces of this system have a $Z$ topological classification according to our criterion. This $Z$ topological invariant characterizes the chiral hinge states
of the second-order TI phase and the 
chiral hinge states can be attributed to the nontrivial boundary topology.

The second example is the 2D BBH model \cite{Benalcazar2017a},  the low-energy effective bulk Hamiltonian of which can be written as \cite{Benalcazar2017}
\beqn
H_2^{(2)}=k_1\gamma_1^{(2)}+k_3\gamma_2^{(2)}+m_0\gamma_3^{(2)}+m_1\gamma_{4}^{(2)}.
\eeqn
 $H_2^{(2)}$ respects the chiral symmetry $\mathcal{C}=\prod_{i=1}^{4}\gamma_i^{(2)}$ and  belongs to the AIII symmetry class. It can be checked that $H_2^{(2)}$ and its double do not allow SPEMT, which indicates that the edges of this system have a $Z$ topological classification according to the given criterion.
This is consistent with the fact that the 1D boundaries of the BBH model are topologically nontrivial and described by the two-band SSH model \cite{Luo2024,Luo2023a}.

\begin{table*}[htb]
\centering
\setlength\tabcolsep{1.5pt}
\renewcommand{\multirowsetup}{\centering}
\renewcommand{\arraystretch}{1.1}
\caption{ Some typical examples of boundary TIs and TSCs described by ${H}_{d}^{(n+1)}$. In columns $2-7$, we list the different representations of the anti-commuting Gamma matrices $\gamma_{1,\cdots,6}^{(g)}$, which give rise to the models ${H}_{1,\cdots,9}$ displaying higher-order topology. The Pauli matrices $\rho$, $\sigma$, $s$, and $\tau$ act in the orbital, orbital, spin, and particle-hole space, respectively. The symbols $\bm{\times}$ in the sixth and seventh columns denote the absence of corresponding Gamma matrices in the Hamiltonian $H_d^{(n+1)}$. In the eighth column, we list the dimension of the system. In the ninth column, we list the type of boundary TIs or TSCs. In the tenth column, we list the order of the HOTPs described by ${H}_{d}^{(n+1)}$. In the eleventh column, we list the type of the gapless boundary states of ${H}_{d}^{(n+1)}$. In the twelfth column, we list the symmetry class of the system, specified by the $\mathcal{T}$, $\mathcal{P}$, and $\mathcal{C}$ symmetries in Table~\ref{tabsy}.}
\begin{tabular}{|c|c|c|c|c|c|c|c|c|c|c|c|}
\hline
Models&$\gamma_1^{(g)}$&$\gamma_2^{(g)}$ & $\gamma_3^{(g)} $ & $\gamma_4^{(g)} $ & $\gamma_5^{(g)} $&$\gamma_6^{(g)}$ &dimensions ($d$)& TIs or TSCs &orders ($n+1$)&boundary states&AZ classes\\
\hline
${H}_1$&$\sigma_3s_1$&$\sigma_3s_2$ &$\sigma_1s_0$ & $\sigma_2s_0$ &$\bm{\times}$&$\bm{\times}$ &2D&edge TI&second-order&corner states &AIII\\
\hline
${H}_2$&$\tau_1s_1$&$\tau_1s_2$&$\tau_3s_0$&$\tau_2s_0$&$\bm{\times}$&$\bm{\times}$ &2D&edge TSC&second-order& corner states &BDI\\
\hline
${H}_3$&$\tau_3\sigma_1s_3$&$\tau_3\sigma_2s_0$&$\tau_3\sigma_3s_0$&$\tau_1\sigma_0s_0$&$\bm{\times}$&$\bm{\times}$ &2D&edge TSC&second-order& corner states&DIII \\
\hline
${H}_4$&$\sigma_1s_1$&$\sigma_1s_2$&$\sigma_1s_3$&$\sigma_3s_0$&$\sigma_2s_0$&$\bm{\times}$ &3D&surface TI&second-order&chiral hinge states &A\\
\hline
${H}_5$&$\rho_0\sigma_1s_1$&$\rho_0\sigma_1s_2$&$\rho_0\sigma_1s_3$&$\rho_0\sigma_3s_0$&$\rho_2\sigma_2s_0$&$\bm{\times}$ &3D&surface TI&second-order&helical hinge states &AII\\
\hline
${H}_6$&$\tau_1s_1$&$\tau_1s_2$&$\tau_1s_3$&$\tau_3s_0$&$\tau_2s_0$&$\bm{\times}$ &3D&surface TSC&second-order&chiral hinge states &D\\
\hline
${H}_7$&$\tau_3\sigma_1s_1$&$\tau_3\sigma_1s_2$&$\tau_3\sigma_1s_3$&$\tau_1\sigma_0s_0$&$\tau_3\sigma_3s_0$&$\bm{\times}$ &3D&surface TSC&second-order&helical hinge states &DIII\\
\hline
${H}_8$&$\rho_3\sigma_1s_1$&$\rho_3\sigma_1s_2$&$\rho_3\sigma_1s_3$&$\rho_3\sigma_2s_0$&$\rho_1\sigma_0s_0$&$\rho_2\sigma_0s_0$ &3D&hinge TI&third-order&corner states &AIII\\
\hline
${H}_9$&$\tau_3\sigma_1s_1$&$\tau_3\sigma_1s_2$&$\tau_3\sigma_1s_3$&$\tau_3\sigma_3s_0$&$\tau_1\sigma_0s_0$&$\tau_2\sigma_0s_0$ &3D&hinge TSC&third-order&corner states &BDI\\
\hline
\end{tabular}
\label{tab4}
\end{table*}

%\hline
%size($H_{q+1}^{2}$)&$2^{2}$&$2^{2}$ & $2^{3}$ &$2^{3}$ &$2^{2}$&$2^{2}$&$2^{2}$&$2^{3}$&$2^{3}$&$2^{3}$&$2^{3}$&$2^{3}$&$2^{3}$&$2^{4}$&$2^{4}$\\
%\hline
%size($H_{q+1}^{3}$)&$2^{3}$&$2^{3}$ & $2^{4}$ &$2^{4}$ &$2^{3}$&$2^{3}$&$2^{3}$&$2^{4}$&$2^{4}$&$2^{4}$&$2^{4}$&$2^{4}$&$2^{3}$&$2^{5}$&$2^{5}$\\

\subsection{Nontrivial boundary topology at arbitrary dimensions}
\label{IIIb}
We now further demonstrate that the nontrivial topology can arise at the $(d-n)$D boundary for a $d$D Dirac Hamiltonian with ($n+1$) Dirac mass terms with $0<n<d$. Generally, we consider the Hamiltonian
\beqn
{H}_d^{(n+1)}=\sum_{i=1}^{d}k_i\gamma_i^{(g)}+\sum_{j=0}^{n}m_j\gamma_{d+j+1}^{(g)},
\label{gd}
\eeqn
where Gamma matrices $\gamma_j^{(g)}$ form the Clifford algebra and satisfy the anti-commutation relation $\{\gamma_j^{(g)},\gamma_{j^{\prime}}^{(g)}\}=2\delta_{jj^{\prime}}I^{(g)}$, with $j,j^{\prime}=0,\cdots d+n+1$ and $d+n\leq 2g$. For the system described by ${H}_d^{(n+1)}$, its bulk and $(d-1)$D,$\cdots$, $(d-n+1)$D boundaries are topologically trivial as the corresponding Hamiltonians involve more than one Dirac mass term. However, we find that 
 the $(d-n)$D boundaries of this system can possess nontrivial topology if ${H}_d^{(n+1)}$ does not allow SPEMT.

Surface projection theory provides a powerful tool for understanding and revealing nontrivial boundary topology \cite{Khalaf2018,Khalaf2018a}. 
In the following, we conduct boundary projection analysis to extract boundary Hamiltonians.  Under the open boundary conditions along the $r_j$ directions for $j=d-n+1,\cdots,d$, ${H}_d^{(n+1)}$ can be separated into two parts
\beqn
&&{H}_d^{(n+1)}={H}_q^{(1)}({\bm k_{q}})+{H}_n^{(n)}({\bm r_{n}}),\nonumber\\
&&{H}_q^{(1)}({\bm k_{q}})=\sum_{i=1}^{q}k_i\gamma_i^{(g)}+m_0\gamma_{d+1}^{(g)},\nonumber\\
&&{H}_n^{(n)}({\bm r_{n}})=\sum_{j=q+1}^{d}-i\partial_{r_j}\gamma_j^{(g)}+m_{j-q}\gamma_{j+n+1}^{(g)},
\label{hd}
\eeqn
where we have defined $\bm k_{q}=(k_1,\cdots,k_{q})$ and $\bm {r}_{n}=(r_{q+1},\cdots,r_d)$ with $q=d-n$. To yield zero-energy states, we consider $n$D mass domain walls at $\bm r_n=0$, namely,  $m_{j-q}(r_j)=m_{j-q}>0$ and $m_{j-q}(r_j)=-m_j<0$ for $r_j>0$ and $r_j<0$, respectively, where $m_{j-q}$ is only a function of $r_j$ (we note that the case $n=2$ was discussed in the work \cite{Benalcazar2017}). For such a configuration,
we can show that ${H}_n^{(n)}(\bm r_n)$ hosts zero-energy boundary states  (see Appendix \ref{Appendix B})
\beqn
\Psi_{-,\cdots,-}(\bm r_n)=\mathcal{N}\prod_{j=q+1}^{d}e^{-\int_{0}^{r_j}m_{j-q}(r_j^{\prime})dr_j^{\prime}}\psi_{-,\cdots,-},
\label{wf1}
\eeqn
where $\psi_{-,\cdots,-}$ is the common eigenstates of $\{\mathcal{C}_{q+1},\cdots,\mathcal{C}_j,\cdots,\mathcal{C}_d\}$ with eigenvalue $-1$, namely 
\beqn
\mathcal{C}_j\psi_{-,\cdots,-}=-\psi_{-,\cdots,-},
\eeqn
with $\mathcal{C}_j=i\gamma_j^{(g)}\gamma_{j+n+1}^{(g)}$.
The subspace expanded by  $\Psi_{-,\cdots,-}(\bm r_n)$ can be defined by the projection operator $P_{-,\cdots,-}=\prod_{j=q+1}^{d}P_{-}^{j}$, with $P_{-}^{j}=(1-\mathcal{C}_j)/2$. Thus, the boundary Hamiltonian can be extracted as
\beqn
\tilde{{H}}_q^{(1)}&=&P_{-,\cdots,-}{H}_q^{(1)}P_{-,\cdots,-}\nonumber\\
&=&\sum_{i=1}^{q}k_j\tilde{\gamma}_j^{(g-n)}+m_0\tilde{\gamma}_{d+1}^{(g-n)},
\label{bdh}
\eeqn
where we have defined $\tilde{\gamma}_j^{(g-n)}=P_{-,\cdots,-}\gamma_jP_{-,\cdots,-}$, which are $2^{(g-n)}\times 2^{(g-n)}$ matrix when picking the nonzero block part.

We find that $\tilde{{H}}_q^{(1)}$ has the identical form as the Dirac Hamiltonian ${H}_d^{(1)}$ with dimension reduction. Therefore, boundary phases described by $\tilde{{H}}_q^{(1)}$ with $m_0>0$ and $m_0<0$ are topologically distinguished,  provided that $\tilde{{H}}_q^{(1)}$ lacks SPEMT. In Appendix \ref{Appendix C},
we show that the SPEMT $m_{n+1}\tilde{\gamma}_{d+n+2}^{(g-n)}$ for $\tilde{{H}}_q^{(1)}$ and the SPEMT $m_{n+1}\tilde{\gamma}_{d+n+2}^{(g)}$ for ${H}_d^{(n+1)}$ has a one-to-one correspondence. Therefore, we can make another conclusion that if ${H}_d^{(n+1)}$ lacks SPEMT, then the $(d-n)$D boundaries of this system can possess nontrivial topology.
When $n=0$, our results recover the bulk
topological classification of first-order topological TIs and
TSCs.  When $n\geq 1$, this
criterion determines the topological classification of boundaries, which share the same classification as standard tenfold way for bulk states. Therefore, this criterion provides a unified framework for bulk and boundary nontrivial topology.

\section{construction of boundary TIs and TSCs}
\label{IV}
The topological classification of gapped bulk leads to first-order TIs and TSCs of the full AZ symmetry classes, as schematically illustrated in Fig.~\ref{fig1}(a). Correspondingly, the topological classification of a gapped boundary can lead to boundary TIs and TSCs, as schematically illustrated in Fig.~\ref{fig1}(b). In the following, we construct lattice models of
boundary TIs and TSCs.

\subsection{ Lattice model Hamiltonians}
Dirac lattice Hamiltonian ${H}_{d}^{(1)}$ in Eq.~\eqref{Dirac} can generally describe first-order TIs and TSCs. To construct boundary TIs and TSCs, we aim to construct a boundary Dirac lattice Hamiltonian with a similar form of ${H}_{d}^{(1)}$. With this spirit, we construct the model Hamiltonian
\beqn
{H}_d^{(n+1)}(\bm k_d)&=\sum_{i=1}^{d}\sin k_i\gamma_i^{(g)}+M_0(\bm k_q)\gamma_{d+1}^{(g)}\nonumber\\
&+\sum_{j=1}^{n}M_j( k_{q+j})\gamma_{j+d+1}^{(g)},
\label{ho}
\eeqn
where $q=d-n$, $M_0(\bm k_q)=\mathcal{M}+\sum_{i=1}^{q}(1-\cos k_i)$ with $-2<\mathcal{M}<0$, and $M_{j}=(t+\cos k_{q+j})$ with $0<t<1$.

By taking a similar boundary projection analysis as Eq.~\eqref{bdh}, we extract the boundary Hamiltonian. Under the open boundary conditions along the $r_{q+1},\cdots, r_d$ directions, we separate ${H}_d^{(n+1)}$ into two parts
\beqn
&&{H}_d^{(n+1)}(\bm k_q,\bm r_n)={H}_{q}^{(1)}(\bm k_q)+{H}_{n}^{(n)}(\bm r_{n}),\nonumber\\
&&{H}_{q}^{(1)}(\bm k_q)=\sum\limits_{i=1}^{q}\sin k_i\gamma_{i}^{(g)}+M_0(\bm k_q)\gamma_{d+1}^{(g)},\nonumber\\
&&{H}_{n}^{(n)}(\bm r_{n})=\sum_{j=q+1}^{d}h_{j}(r_j),\nonumber\\
&&h_{j}(r_j)=-i\partial_{r_j}\gamma_{j}^{(g)}
+M_{j-q}(-i\partial_{r_j})\gamma_{j+n+1}^{(g)},
\label{BT}
\eeqn
where ${H}_{n}^{(n)}$ can be regarded as the $n$D generalization of the BBH model and the topological properties of it were studied in our recent work \cite{Luo2024}. It can be shown that ${H}_{n}^{(n)}$ hosts $2^g$ zero-energy boundary states (see Appendix \ref{Appendix B})
\beqn
\Psi_{z_{q+1},\cdots,z_d}(\bm r_n)=\prod_{j=q+1}^df_{z_j}^{(j)}(r_j)\psi_{z_{q+1},\cdots,z_d},
\label{wf2}
\eeqn
where spinor $\psi_{z_{q+1},\cdots,z_d}$ is the common eigenstates of $\{\mathcal{C}_{q+1},\cdots,\mathcal{C}_j,\cdots,\mathcal{C}_d\}$ and  satisfies
\beqn
\mathcal{C}_j\psi_{z_{q+1},\cdots,z_d}=z_j\psi_{z_{q+1},\cdots,z_d},
\eeqn
with eigenvalue $z_j=\pm 1$. $f_{z_j}^{(j)}(r_j)$ is the real space wave function part of the zero-energy end states of 
$h_j$. For $z_j=-1$ and $z_j=1$, $f_{-}^{(j)}(r_j)$ and $f_{+}^{(j)}(r_j)$ are localized close to $r_j=0$ and $r_j=L$, respectively, with $L$ being the length of the system along the $r_j$ direction.
The zero-energy subspace expanded by $\Psi_{z_{q+1},\cdots,z_d}(\bm r_n)$ can be defined by the projection operator $P_{z_{q+1}\cdots z_d}=\prod_{j=q+1}^d(1+z_j\mathcal{C}_j)/2$, which will be used to extract the boundary Hamiltonian.

Because of the separability of
${H}_d^{(n+1)}$ into $\bm k_q$ and $\bm r_{n}$ independent parts, ${H}_d^{(n+1)}(\bm k_q, \bm r_n)$ hosts the gapped boundary states which are extended over the whole Brillouin zone expanded by $\bm k_q$. These gapped boundary states are the common eigenstates of  $\{\mathcal{C}_{q+1},\cdots,\mathcal{C}_d,H_{q}^{(1)}\}$ and can be written as
\beqn
&&\Phi(\bm k_q,\bm r_n)=\prod_{j=q+1}^df_{z_j}^{(j)}(r_j)P_{z_{q+1}\cdots z_d}\phi(\bm k_q),\nonumber\\
&&{H}_{q}^{(1)}(\bm k_q)\phi(\bm k_q)=E(\bm k_q)\phi(\bm k_q).
\label{dl}
\eeqn
It can be exactly shown that
\beqn
{H}_d^{(n+1)}(\bm k_q, \bm r_n)\Phi(\bm k_q,\bm r_n)=E(\bm k_q)\Phi(\bm k_q,\bm r_n).
\eeqn
Thus, the boundary states of ${H}_d^{(n)}(\bm k_q, \bm r_n)$ have the same energy spectra as those of ${H}_{q}^{(1)}(\bm k_q)$. The corresponding boundary Hamiltonian can be extracted by projecting ${H}_{q}^{(1)}$ onto the zero-energy subspace defined by $P_{z_{m+1}\cdots z_d}$, which gives rise to
\beqn
\tilde{{H}}_{q}^{(1)}(\bm k_q)=\sum\limits_{j=1}^{q}\sin k_j\tilde{\gamma}_{j}^{(g-n)}+M_0(\bm k_q)\tilde{\gamma}_{d+1}^{(g-n)},
\label{BD}
\eeqn
where $\tilde{\gamma}_{d+1,j}^{(g-n)}=P_{z_{q+1}\cdots z_d}\gamma_{d+1,j}^{(g)}P_{z_{q+1}\cdots z_d}$.
When picking the non-zero block part, $\tilde{\gamma}_{d+1,j}$ are $2^{(g-n)}\times 2^{(g-n)}$ Gamma matrices.
Thus, we obtain a boundary Dirac lattice Hamiltonian $\tilde{{H}}_{q}^{(1)}(\bm k_q)$, which can generally describe boundary TIs and TSCs of the full AZ symmetry classes.

\begin{table}[htb]
\centering
\setlength\tabcolsep{10pt}
\renewcommand{\multirowsetup}{\centering}
\renewcommand{\arraystretch}{1.2}
\caption{Representation of $\mathcal{T}$, $\mathcal{P}$, and $\mathcal{C}$ symmetries for models $H_{1,...,9}$. The symbol $\times$ denotes the absence of the corresponding symmetry and $K$ is the complex conjugation operator.}
\begin{tabular}{|c|c|c|c|c|}
\hline
Models&AZ classes&$\mathcal{T}$ &$\mathcal{P}$ & $\mathcal{C}$ \\
\hline
$H_1$&AIII&$\times$&$\times$&$\sigma_3s_3$\\
\hline
$H_2$&BDI&$\tau_3s_1 K$&$\tau_2s_2K$&$\tau_1s_3$\\
\hline
$H_3$&DIII&$is_2 K$&$\tau_2s_2K$&$\tau_2$\\
\hline
$H_4$&A&$\times$&$\times$&$\times$\\
\hline
$H_5$&AII&$is_2K$&$\times$&$\times$\\
\hline
$H_6$&D&$\times$&$\tau_2s_2K$&$\times$\\
\hline
$H_7$&DIII&$is_2K$&$\tau_2s_2K$&$\tau_2$\\
\hline
$H_8$&AIII&$\times$&$\times$&$\rho_3\sigma_3s_0$\\
\hline
$H_9$&BDI&$\tau_1\sigma_2s_2K$&$\tau_2s_2K$&$\tau_3\sigma_2s_0$\\
\hline
\end{tabular}
\label{tabsy}
\end{table}

\subsection{Remarks}
For the constructed Hamiltonian $H_d^{(n+1)}$, we highlight several key points. The criterion for the existence of nontrivial boundary topology also applies to the lattice Hamiltonian $H_d^{(n+1)}$. For a given symmetry class, if $H_d^{(n+1)}$ ($\tilde{H}_{q}^{1}$) does not allow SPEMT, a boundary TI or TSC is realized ($n>0$).  When $n=0$, it is known that the condition of the absence of SPEMT for $H_{d}^{(1)}$ ($\tilde{H}_{q}^{1}$) leads to five types out of ten symmetry classes of bulk TIs and TSCs in arbitrary dimensions \cite{Schnyder2008,Kitaev2009,Ryu2010,Chiu2016}. 
Similarly, there are five classes of boundary TIs and TSCs in arbitrary dimensions when $n>0$ given the correspondence between $H_d^{(n+1)}$ and $\tilde{H}_{q}^{1}$. On the other hand, the matrix dimension of $H_{d}^{(1)}$ for describing bulk TIs and TSCs under the given spatial dimension and symmetry class can be known from the extended Clifford algebra \cite{Chiu2014a}. Given the correspondence between  ${H}_{d}^{(n+1)}$ and $\tilde{H}_{q}^{(1)}$, we list the minimal matrix dimension  of ${H}_{d}^{(n+1)}$ for describing boundary TIs and TSCs in different symmetry class with $q=1,2,3$ in Table~\ref{tabtf}. Generally, the matrix dimension of $H_d^{(n+1)}$ is $2^{n+\tilde{g}}$ for describing a TI and TSC at a $(d-n)$D  boundary with $2^{\tilde{g}}$ being the matrix dimension of $\tilde{H}_{q}^{1}$ for describing a first-order TI and TSC.

When a boundary TI or TSC is realized, applying open boundary condition along one direction for $\tilde{H}_q^{(1)} $, there are robust gapless boundary states at the $(d-n-1)$D boundaries, which is a hallmark of a $(n+1)$th order topological phase. Thus, our constructions facilitate the realization of higher-order TIs and TSCs in arbitrary dimensions with arbitrary orders. By expanding $M_0(\bm k_q)$ at $\bm k_q=0$,  the analytical solution of the gapless boundary states of ${H}_d^{(n+1)}$ can be  derived (see Appendix \ref{Appendix B}). These gapless boundary states are robust as long as the bulk and boundary energy gaps persist. Thus, we can make deformation to $H_{d}^{(n+1)}$ by extending the masses $M_{0,\cdots, n}$ to generally involve $k_1,\cdots,k_d$ while maintaining the bulk and boundary energy gaps. Under this deformation, the gapless boundary states still exist and the HOTPs are not trivialized, but the lattice Hamiltonian on $(d-n)$D boundaries may not be well-defined.

In ${H}_d^{(n+1)}$, we require that $M_{j-q}$ is only a function of $k_{j}$ with $j=q+1,\cdots,d$ and $M_0$ is only a function of $\bm k_q$. We emphasize that these conditions for realizing a boundary TI or TSC can be further relaxed, allowing $M_{j}$ to be a general function of $\bm k_d$. In this case, provided that ${H}_{n}^{(n)}(\bm k_q,\bm r_n)$ behaves as an $n$th order TI over all the values of $\bm{k}_q$ in the Brillouin zone, we can project ${H}_{q}^{(1)}$ onto the subspace spanned by the zero-energy boundary states of ${H}_{n}^{(n)}$. This projection process results in a boundary Dirac lattice Hamiltonian of a TI or TSC.
This analysis applies to the models studied in the work \cite{Li2021a}, which proposed to realize time-reversal invariant boundary TSCs.

Furthermore, the anti-commutation relations among the matrices in $H_d^{(n+1)}$ can also be relaxed to achieve boundary TIs and TSCs. For instance, by setting $d = 2$ and $n = 1$, and by removing the anti-commuting relations restrictions on $\{\gamma_{1}^{(g)},\gamma_{2}^{(g)},\gamma_{3}^{(g)},\gamma_{4}^{(g)}\}$,  the 2D SSH and 2D crossed SSH models can be realized as discussed in the work  \cite{Luo2023a}. Similar to the 2D BBH model, the edges of these two models can also exhibit nontrivial topology described by the two-band SSH model. This suggests that relaxing the anti-commutation relations among the matrices in $H_d^{(n+1)}$ opens up extensive possibilities for realizing boundary TIs and TSCs.

%For a given symmetry class and dimension $d$, the matrix dimension of $H_d^{(n+1)}$ for describing  a boundary TI or TSC is $2^n\tilde{g}$. Here $\tilde{g}$ is the matrix dimension of the boundary Dirac Hamiltonian $\tilde{H}_{q}^{(1)}$, which describes a first-order TI or TSC 

\section{Typical examples in 2D and 3D}
\label{V}
Building on the matrix dimension correspondence given in the Table~\ref{tabtf}, we can obtain boundary TIs and TSCs of the full AZ symmetry classes (see Appendix~\ref{appendix e}). In Table~\ref{tab4}, we present some typical examples of boundary TIs or TSCs in 2D and 3D. In the following, we explain these examples case by case.

%further consider the concrete matrix representation of $H_{d}^{(n+1)}$ with $q=1,2$ in Table~\ref{tab4}, which 

%$H_{q+1}^{(1)}$ and $H_{q+2}^{(3)}$ with $q=1$,  $H_{q+1}^{(1)}$ with $q=2$  under certain symmetry class with  in Table~\ref{tab4}, 

%It is known that for a gapped system in arbitrary dimensions, there are five out of ten symmetry classes with nontrivial topological classification, corresponding to five types of TIs or TSCs \cite{Schnyder2008,Kitaev2009,Ryu2010,Chiu2016}. Similarly, there are five classes of boundary TIs and TSCs in arbitrary dimensions.In Table~\ref{tab4}, we present some typical examples of boundary TIs or TSCs in 2D and 3D by considering the concrete representation of the Gamma matrices. By considering the concrete matrix representation of $H_{2}^{2}$, $H_{3}^{2}$, and $H_{3}^{3}$  under certain symmetry class, we present some typical examples of boundary TIs or TSCs in 2D and 3D in Table~\ref{tab4}.

\subsection{Edge, surface, and hinge TIs/TSCs}

\begin{figure}
\centering
\includegraphics[width=3.5in]{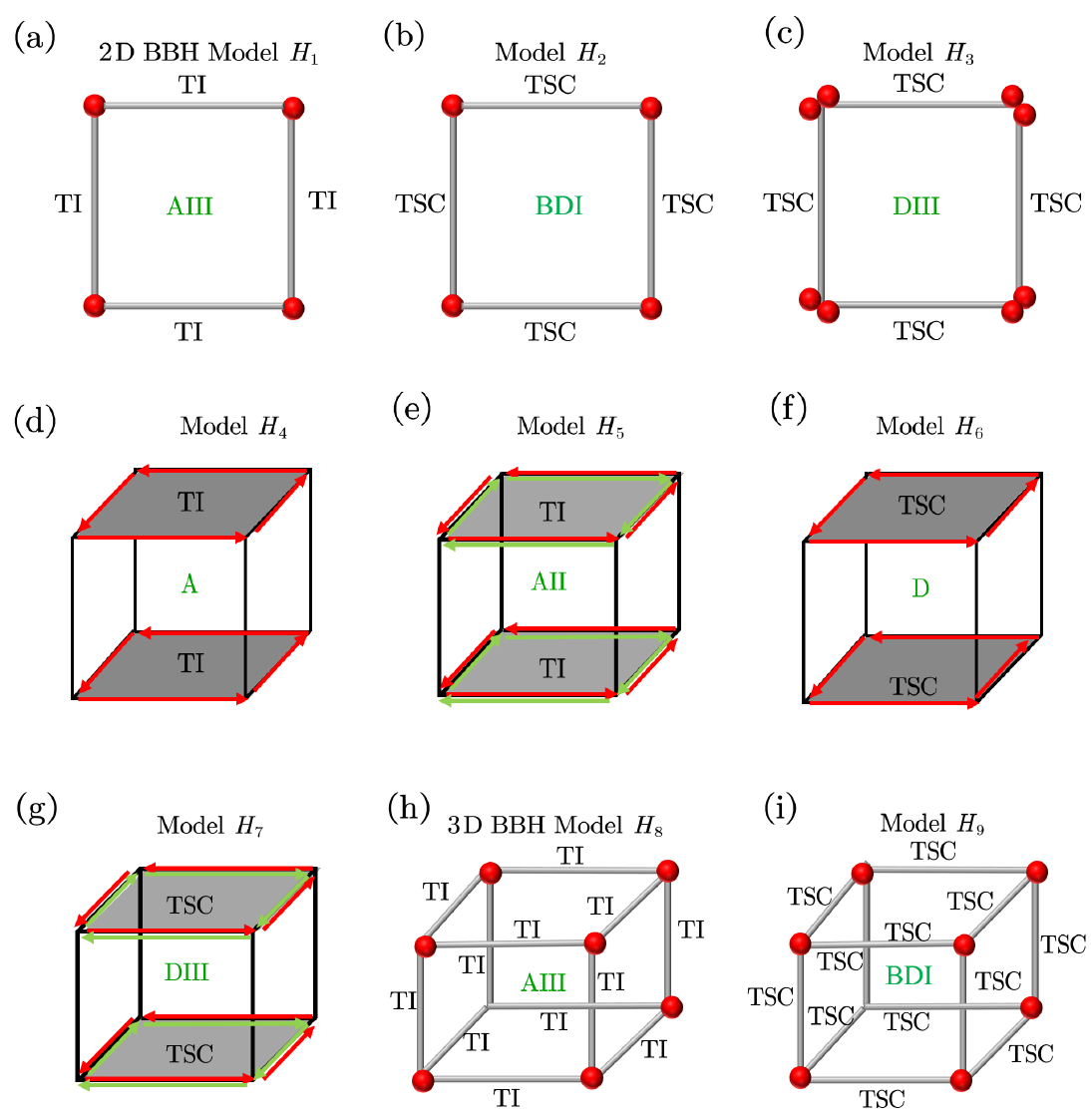}
\caption{The schematic illustration of the boundary TIs or TSCs described by models $H_{1,\cdots,9}$ presented in Table~\ref{tab4}. The gray boundaries, including the edges, surfaces, and hinges are topologically nontrivial. The red dots and lines with arrows denote the gapless corner states and hinge states, respectively. The green letters label the symmetry class of both the bulk and boundaries.}
\label{fig2}
\end{figure}

For 1D gapped systems,  there are nontrivial topological classifications for AIII, BDI, D, DIII, and CII symmetry classes, corresponding to the realization of 1D TIs or TSCs \cite{Schnyder2008,Kitaev2009,Ryu2010,Chiu2016}.  For example, a 1D TI in AIII symmetry class can be realized by the SSH model; a 1D TSC in the BDI symmetry class can be realized in the Kitaev chain \cite{Kitaev2001}; a 1D TSC in the DIII symmetry can be realized by considering time-reversal symmetric $p$-wave superconducting pairing.

We now consider 2D systems with edge TIs and TSCs described by $H_{d}^{(n+1)}$ in the AIII, BDI, and DIII symmetry classes, respectively. By setting $d=2$ and $q=n=1$ in Eq.~\eqref{ho}, the bulk and boundary Hamiltonians ${H}_{d}^{(n+1)}$ and $\tilde{H}_q^{(1)}$ are, respectively, given by
\beqn
&&{H}_{2}^{(2)}=\sum_{i=1}^{2}\sin k_i\gamma_i^{(g)}+\sum_{j=0}^{1}M_j(k_{j+1})\gamma_{j+3}^{(g)},\nonumber\\
&&\tilde{H}_{1}^{(1)}=\sin k_1\tilde{\gamma}_1^{(g-1)}+M_0(k_1)\tilde{\gamma}_3^{(g-1)}.
\label{2DSSH}
\eeqn
Here $g\geq 2$ because $d+n \leq 2g $. For ${H}_{2}^{(2)}$,  we present three representations of the gamma matrices, detailed in rows 2-4 of Table~\ref{tab4}, corresponding to models ${H}_1$, ${H}_2$, and ${H}_3$, respectively.
The $4\times 4$ model ${H}_1$ associated with $g=2$ is exactly the 2D BBH model \cite{Benalcazar2017a,Luo2023a}, whose edge Hamiltonian $\tilde{H}_{1}^{(1)}$ behaves as a two-band SSH model. Because of the equivalent roles of different directions in ${H}_{2}^{(2)}$, all the edges of the BBH model are a 1D TI
 in the AIII symmetry class \cite{Luo2023a}, as schematically illustrated in Fig.~\ref{fig2}(a). ${H}_2$  serves as a superconducting generalization of the 2D BBH model, which hosts Majorana corner states. This model was studied in references \cite{Wang2018a,Tiwari2020a} and all the edges of this model are a 1D TSC in the BDI symmetry class, as schematically illustrated in Fig.~\ref{fig2}(b). Distinguished from ${H}_1$ and ${H}_2$,  ${H}_3$ is $8\times 8$ associated with $g=3$, which hosts Kramers pairs of Majorana corner states. For this model, all the edges are a time-reversal invariant TSC in the DIII symmetry class, as schematically illustrated in Fig.~\ref{fig2}(c). For models ${H}_{1,2,3}$, we emphasize that the corner states are simultaneous eigenstates of both edge Hamiltonians along the $r_1$ and $r_2$ directions, as revealed in the references \cite{Benalcazar2017,Luo2024,Luo2023a}.

In 2D systems,  gapped bulk states have a nontrivial topological classification
 in the A, D, DIII, AII, and C symmetry classes \cite{Schnyder2008,Kitaev2009,Ryu2010,Chiu2016}. For example, a 2D TI in the A symmetry class can be realized by the Qi-Wu-Zhang model \cite{Qi2006}; a 2D TI in the AII symmetry class can be realized by the Bernevig-Hughes-Zhang model \cite{Bernevig2006}; a 2D TSC in the D symmetry class can be realized in the $p_x+ip_y$ superconductor 
 \cite{Bernevig2013}; a 2D TSC in the DIII symmetry class can be realized in the time-reversal symmetric $p_x\pm ip_y$ superconductor \cite{Bernevig2013}.

Here, we present the corresponding surface TIs and TSCs in the A, AII, D, and DIII symmetry classes in 3D systems, respectively. By setting $d=3, q=2$, and $n=1$ in Eq.~\eqref{ho}, the bulk and boundary Hamiltonians ${H}_{d}^{(n+1)}$ and $\tilde{H}_q^{(1)}$ are, respectively, specified by
\beqn
&&{H}_{3}^{(2)}=\sum_{i=1}^{3}\sin k_i\gamma_i^{(g)}+M_0(\bm k_{2})\gamma_{4}^{(g)}+M_1(k_{3})\gamma_{5}^{(g)},\nonumber\\
&&\tilde{H}_{2}^{(1)}=\sum_{i=1,2}\sin k_i\tilde{\gamma}_i^{(g-1)}+M_0(\bm k_2)\tilde{\gamma}_4^{(g-1)},
\label{32}
\eeqn
where $g\geq 2$. For ${H}_{3}^{(2)}$, we present four representations of the gamma matrices, detailed in rows 5-8 of the Table~\ref{tab4}, corresponding to models ${H}_4$, ${H}_5$, ${H}_6$, and ${H}_7$, respectively. In ${H}_4$ (A symmetry class) and ${H}_5$ (AII symmetry class), we take $g=2$ and $g=3$, respectively.  We note that $H_4$ is discussed in the work \cite{das2023hybrid}. The boundary Dirac Hamiltonians $\tilde{H}_{2}^{(1)}$ for ${H}_4$ and ${H}_5$ are $2\times 2$ and $4\times 4$, respectively. Thus, the $(001)$ surfaces for ${H}_4$ and ${H}_5$ behave as a surface Chern insulator and surface time-reversal invariant TI, respectively, as schematically illustrated in Figs.~\ref{fig2}(d) and \ref{fig2}(e). Similarly, for  ${H}_6$ (D symmetry class) and ${H}_7$ (DIII symmetry class), the $(001)$ surfaces of which behave as a TSC and time-reversal invariant TSC, respectively, as schematically illustrated in Figs.~\ref{fig2}(f) and \ref{fig2}(g).

\begin{figure}
\centering
\includegraphics[width=3.5in]{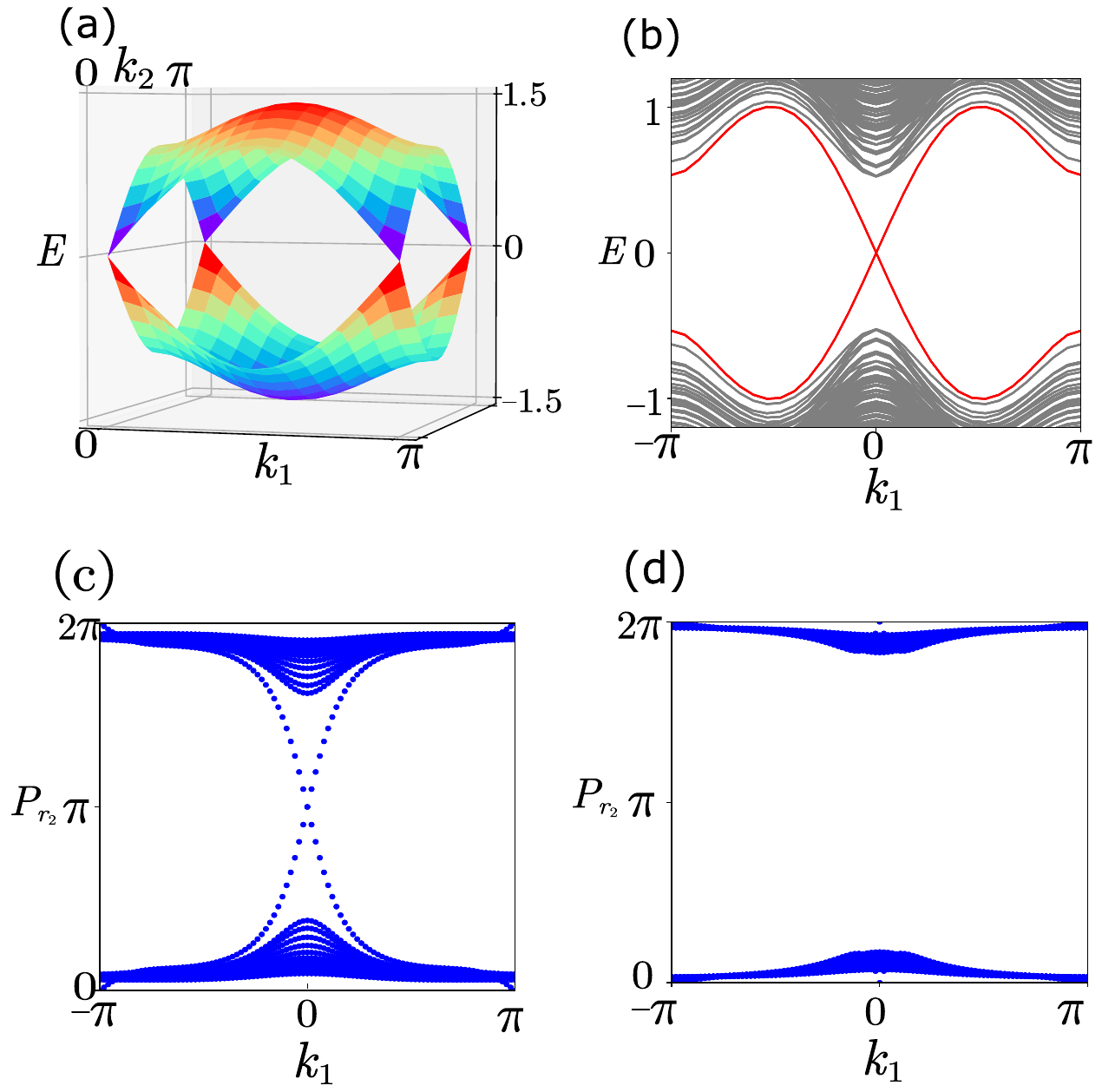}
\caption{(a) The energy spectrum of the four gapless surface Dirac cones on the (001) surface of the normal states of $H_7$. (b) The energy spectrum of $H_7$ in a wire geometry along $k_{1}$ with
open boundary conditions on both $r_2$ and $r_3$ directions. The in-gap dispersions are of four-fold degeneracy, corresponding to four helical Majorana modes located separately at four hinges. (c) The Wannier spectrum $P_{r_2}$ obtained by performing the Wilson loop along $k_2$ for a slab geometry on the
(001) surface. In (a),  we take $t=0.5$. In (b)-(c), we take $\Delta_0=-\Delta_1=t=0.5,\mu=0$. In (d) we take $t=0.5,\mu=0,\Delta_0=1.2,\Delta_1=-0.5$. }
\label{fig3}
\end{figure}

A 3D system also has the $(d-2)$D boundaries, namely 1D hinges. We focus on hinge TIs or TSCs by choosing $d=3, q=1$, and $n=2$ in Eq.~\eqref{ho}, then bulk and boundary Hamiltonians ${H}_{d}^{(n+1)}$ and $\tilde{H}_q^{(1)}$ are, respectively, given by
\beqn
&&{H}_{3}^{(3)}=\sum_{i=1}^{3}\sin k_i\gamma_i^{(g)}+\sum_{j=0}^{2}M_j(k_{j+1})\gamma_{j+4}^{(g)},\nonumber\\
&&\tilde{H}_{1}^{(1)}=\sin k_1\tilde{\gamma}_1^{(g-2)}+M_0(k_1)\tilde{\gamma}_4^{(g-2)},
\eeqn
where $g\geq 3$. For ${H}_{3}^{(3)}$, we consider two representations of the Gamma matrices by setting $g=3$, which are listed in rows 9-10 of Table~\ref{tab4}, corresponding to models ${H}_8$ and ${H}_{9}$, respectively. ${H}_8$ is exactly the 3D BBH model \cite{Benalcazar2017,Benalcazar2017a,Luo2024}, all hinges of which behave as a 1D TI  described by the SSH model, as schematically illustrated in Fig.~\ref{fig2}(h). ${H}_{9}$ is the superconducting generalization of the 3D BBH model. This model hosts Majorana corner states and all the hinges are a 1D TSC in the BDI symmetry class, as schematically illustrated in Fig.~\ref{fig2}(i).  For the given symmetry class defined by the local symmetries (see Tables \ref{tab4} and \ref{tabsy}), it can be examined that $H_{1,\cdots,9}$ do not allow SPEMT, which is consistent with the presence of nontrivial boundary topology.

\subsection{Surface time-reversal invariant TSC described by ${H}_7$ }

For concreteness, we further take the model $H_7$ as an example to reveal
the nontrivial boundary topology.
The theoretical model ${H}_7$ is associated with the Bogoliubov-de Gennes Hamiltonian, ${H}_{\rm BdG}(\bm k)=\sum_{\bm k}\psi_{\bm k}^{\dagger}H_{7}(\bm k)\psi_{\bm k}$, where
\beqn
H_{7}(\bm k)=\begin{pmatrix}
\label{TI}
{H}_{\text{N}}( \bm{k}) -\mu& \Delta(\bm{k_{||}})\\
\Delta^{\dagger}(\bm{k}_{||})&-{H}_{\text{N}}(\bm{k})+\mu\end{pmatrix},
\eeqn
and $\psi_{\bm k}=(c_{\bm{k} \uparrow}^{a},c_{\bm{k}, \downarrow}^{a},c_{\bm{k} \uparrow}^{b},c_{\bm{k}\downarrow}^{b},-c_{-\bm{k} \downarrow}^{a\dagger},c_{-\bm{k} \uparrow}^{a\dagger}, -c_{-\bm{k} \downarrow}^{b\dagger},c_{-\bm{k} \uparrow}^{b\dagger})$.
Here $a$ and $b$ label the different orbitals. $\uparrow$ and $ \downarrow$ denote the spin up and down, respectively. The normal-state Hamiltonian is ${H}_{\text{N}}( \bm{k})=M_1(k_3)\sigma_zs_0+\sum_{i=1}^{3}\sin k_i\sigma_1s_i$, where $M_1(k_3)=(t+\cos k_3) $ with model parameters $0<t<1$. $\mu$ is the chemical potential, and we take $\mu=0$ in Eq.~\eqref{32}. $\Delta(\bm k_{||})=\Delta_0+\Delta_1(\cos k_1+\cos k_2)$ denotes the  $s_{\pm}$-wave superconducting  order parameter with $\bm k_{||}$ being the in-plane momentum vector. $\Delta(\bm k_{||})$ is associated with $M_0(\bm k_{2})$ in Eq.~\eqref{32}. $H_{7}(\bm k)$ respects the time-reversal and particle-hole symmetries, given by ${\mathcal{T}}=is_2K$ and ${\mathcal{P}}=\tau_2s_2K$, respectively, and therefore belongs to the DIII symmetry class.

It can be verified that $H_{\text{N}}$ has band inversion at the four time-reversal invariant points  $(\alpha_1,\alpha_2,\pi)$ with $\alpha_{1,2}=0,\pi$,  indicating that it is topologically trivial according to the Fu-Kane index \cite{Fu2007}. Despite  being topologically trivial, $H_{\text{N}}$ hosts four surface Dirac cones on the $(001)$ surface, as shown in Fig.~\ref{fig3}(a). When further considering the $s_{\pm}$-wave superconducting pairing, the four Dirac cones are gapped. The corresponding Hamiltonian for the (001) surface can be explicitly written as (see Appendix \ref{Appendix D})
\beqn
\tilde{{H}}_{2}^{(1)}=-\Delta(\bm k_{||})\tilde{\tau}_y\tilde{s}_z-(\sin k_x\tilde{s}_y-\sin k_y\tilde{s}_x),
\label{bdh1}
\eeqn
where Pauli matrices $\tilde{\tau},\tilde{s}$ act on the surface states subspace. The time-reversal and particle-hole symmetries of $\tilde{{H}}_{2}^{(1)}$ are given by $\tilde{\mathcal{T}}=i\tilde{s}_yK$ and $\tilde{\mathcal{P}}=-i\tilde{\tau}_x\tilde{s}_xK$, respectively. Obviously, $\tilde{{H}}_{2}^{(1)}$ is block-diagonal in the eigenbasis of $\tilde{\tau}_y$. Especially, the two $2\times 2$ block Hamiltonians, related by the time-reversal symmetry, behave as a 2D chiral topological superconductor when $|\Delta_0|<2|\Delta_1|$. Thus, $\tilde{{H}}_{2}$ behaves as a time-reversal invariant TSC, giving rise to the helical Majorana hinge modes, as shown in Fig.~\ref{fig3}(b).
  
This surface TSC can be characterized by the surface topological invariant defined on a slab geometry. In Fig.~\ref{fig3}(c), we present the Wannier bands $P_{r_2}$ obtained by performing the Wilson loop along $k_2$ for a slab geometry on the
(001) surface. This Wannier spectrum exhibits a winding with the evolution of $k_1$, which explicitly characterizes the gapless helical Majorana hinge modes in Fig.~\ref{fig3}(b). For a comparison, we present the Wannier spectrum $P_{r_2}$ for a topologically trivial case in Fig.~\ref{fig3}(d), which exhibits no winding with the evolution of $k_1$. 

We emphasize that the surface TSC realized here is closely related to the idea of turning a topological insulator to a topological superconductor by the superconducting proximity effect \cite{Fu2008, Wang2018b}. Especially, Majorana helical hinge states can also be realized in iron-based superconductors when considering $s_{\pm}$ pairing \cite{Zhang2019}. The distinct feature here is that the realized surface TSC is well defined as it can be described by a lattice Hamiltonian across the full surface Brillouin zone.

\section{Discussion and Conclusion}
\label{VI}

The classification of boundary TIs and TSCs aligns with the standard tenfold way classification for gapped bulk systems. Similarly, the classification of defects, as established in Ref.~\onlinecite{Teo2010}, follows the tenfold way by replacing the dimension with codimension of defects. Additionally, the classification of extrinsic  $n$th-order TIs and TSCs in a $d$D system has the same classification of codimension $n$ topological defects, as noted in Refs.~\onlinecite{Langbehn2017,Geier2018}. Consequently, the classifications of boundary topological phases, topological defects, and extrinsic higher-order topological phases share the same foundational framework. While our work builds upon this well-established framework, its novelty and significance lie in the systematic study of boundary TIs and TSCs by providing model construction and unified criterion for their emergence within Dirac modes, offering new sights for higher-order topological phases and their unified construction in arbitrary dimensions of arbitrary orders.

%A theoretical work predicted that

Boundary TIs can potentially be realized in van der Waals topological materials. A theoretical study \cite{Yoon20201} demonstrated that the $\alpha$ phase of Bi$_4$I$_4$ functions as a higher-order TI, hosting helical hinge states. This behavior can be understood as follows: the monolayer $\alpha$-Bi$_4$I$_4$ acts as a 2D TI, while the primitive unit cell of the 3D bulk material includes two layers. The different intercell and intracell tunneling along the $z$ direction couples the 2D TI in a dimerized manner, which leads to the emergence of a surface TI. Some signatures of this physics were observed in an experiment \cite{Huang2021a}. Similar physics may occur in van der Waals topological materials 1T$^{\prime}$-WTe$_2$ \cite{Xu2023}.

Topological states can be generalized to gapless systems, such as Dirac and Weyl semimetals \cite{Armitage2018}. For some systems hosting nodal points for the surface Hamiltonians, we can also define boundary topological semimetals which feature hinge flat bands connected by the nodal points. Furthermore,  boundary Floquet TI or TSC  can also be defined.
For example,  the surface Floquet Chern insulator could be realized in a weak TI  with band inversion at $(0,0,0)$ and $(0,0,\pi)$ points
under the irradiation of circularly polarized light. This type of weak TI has two surface Dirac cones which have the same low-energy effective Hamiltonians as monolayer graphene. Since the circularly polarized light can drive graphene into a Floquet Chern insulator \cite{Kitagawa2008}, we can reasonably expect that a surface Floquet Chern insulator can be realized when this type weak TI is irradiated by the circularly polarized light.

Up to now, we have not considered any crystal symmetry and the bulk of the constructed TIs and TSCs is topologically trivial but host gapless boundary states with codimension larger than one, which fall into the category of extrinsic HOTPs or boundary obstructed topological phases \cite{Khalaf2021}. We emphasize that once considering certain crystal symmetry, the bulk of boundary TIs and TSCs can be topologically nontrivial, which drives the system into the category of intrinsic HOTPs. We take the 2D BBH model ($H_1$) as an example. Without the four-fold rotation symmetry ($C_4$), the bulk of $H_1$
is topologically trivial and corner states can be removed by closing the edge states energy gap \cite{Benalcazar2017,Khalaf2021}. Once considering $C_4$ symmetry, the bulk of $H_1$
is topologically nontrivial and the corner states can not be removed without closing the bulk energy gap and breaking $C_4$ symmetry \cite{Benalcazar2017,Khalaf2021}. Therefore, boundary TIs and TSCs can also host nontrivial bulk topology when further involving the crystal symmetry and enable the realization of intrinsic HOTPs.

%\textcolor{red}{It is an interesting question to ask whether other models presented in Table II can be intrinsic HOTPs when crystalline symmetries are considered, which is beyond the scope of this work and left for further study.}

%We now discuss the relations of boundary TIs and TSCs with the intrinsic and extrinsic HOTPs. In our model construction and 

In summary, we systematically study boundary TIs and TSCs of the full AZ symmetry class, 
Through boundary projection analyses of the Dirac continuum model, we demonstrate that nontrivial boundary topology can arise from a system with trivial bulk topology. We provide a unified criterion for the emergence of nontrivial bulk and boundary topology. Additionally, we construct lattice models of realizing boundary TIs and TSCs within the full AZ symmetry class, facilitating the realization of HOTPs in arbitrary dimensions with arbitrary orders. Finally, we present a detailed analysis on some examples of boundary TIs and TSCs in 2D and 3D systems, which offers a unified description of many distinct models.

\section{Acknowledgments}
We thank R. Queiroz for providing valuable comments. This work is supported by the National Natural Science Foundation of China (Grant No. 12274333).

%For the complex Clifford algebra $Cl_{}$, there are $2g$ generators, labelled as $e_{j}$ for $j=1,\cdots,2g$, which satisfy the anti-commutation relation $\{e_{j},e_{j^{\prime}}\}=2\delta_{jj^{\prime}}$. Another operator anti-commuting with the $2g$ generators  can be written as
%\beqn
%e_{2g+1}=i^ne_1e_2,\cdots,e_{2g}, \{e_{2g+1}, %e_j\}=0,e_{2g+1}^2=1.\nonumber\\
%\eeqn$\gamma^{(g)}_{1,2,\cdots,2g+1}$, which are the direct product of $g$ sets of Pauli matrices and satisfy $(\gamma_j^{(g)})^{2}=1,(\gamma_j^{(g)})^{\dagger}=\gamma_j^{(g)}$. 

\appendix
\section{Clifford algebra and Gamma matrices}
\label{Appendix A}
A complex Clifford algebra \( Cl_p \) has \( p \) generators \( e_{1,\cdots,p} \) that satisfy the anti-commutation relation $\{e_i, e_j\} = 2 \delta_{ij}$ for $i,j\in{1,\cdots,p}$.
The algebra \( Cl_p \)  is constructed from all possible products of these generators, \( e_1^{p_1} e_2^{p_2} \cdots e_p^{p_p} \) with \( p_i = 0 \) or \( 1 \), and their linear combinations with complex coefficients. 
The $p$ generators for $p>1$  can be faithfully represented in terms of $p$ anti-commuting $2^g\times 2^g$ Gamma matrices $\gamma^{(g)}_{1,2,\cdots,p}$, which are the direct product of $g$ sets of Pauli matrices and satisfy $(\gamma_j^{(g)})^{2}=1,(\gamma_j^{(g)})^{\dagger}=\gamma_j^{(g)}$. For a given $g$, there are $2g+1$ anti-commuting Gamma matrices $\gamma^{(g)}_{1,2,\cdots,2g+1}$. 
Therefore, $g={[p/2]}$, where $[x]$ equals to the largest integer not exceeding $x$.
When $g$=1, $\gamma_{1,2,3}^{(1)}$ correspond to the three Pauli matrices $\sigma_{1,2,3}$. When $g$=2, $\gamma_{1,2,3,4,5}^{(2)}$ correspond to five anti-commuting Dirac matrices, which can be chosen as
\beqn
\begin{aligned}
&\gamma_{1,2,3}^{(2)}=\sigma_3\otimes \sigma_{1,2,3}, \gamma_4^{(2)}=\sigma_1\otimes \sigma_{0},\\
&\gamma_5^{(2)}=-\gamma_1^{(2)}\gamma_2^{(2)}\gamma_3^{(2)}\gamma_4^{(2)}=\sigma_2\otimes \sigma_0.
\end{aligned}
\eeqn
The anti-commuting $2^g\times 2^g$ Gamma matrices $\gamma^{(g)}_{1,2,\cdots,2g+1}$ can be generically obtained according to the iteration relation
\beqn
\begin{aligned}
&\gamma_{1,2,\cdots,2g-1}^{(g)}=\sigma_3\otimes \gamma_{1,2,\cdots,2g-1}^{(g-1)},\\
& \gamma_{2g}^{(g)}=\sigma_1\otimes I^{(g-1)}, \gamma_{2g+1}^{(g)}=\sigma_2\otimes I^{(g-1)},
\end{aligned}
\eeqn
where $I^{(g-1)}$ denotes the $2^{g-1}\times 2^{g-1}$ identity matrix.

On the other hand, a real Clifford algebra $C l_{p, q}$ has $p+q$ generators satisfying
\beqn
\begin{aligned}
&&\left\{e_i, e_j\right\} & =0(i \neq j), \\
&&e_i^2 & = \begin{cases}-1, & 1 \leqslant i \leqslant p, \\
+1, & p+1 \leqslant i \leqslant p+q.\end{cases}
\end{aligned}
\eeqn
The products of generators and their combinations with real coefficients form a real Clifford algebra $C l_{p, q}$. These generators can be faithfully represented by using $p+q$ anti-commuting Gamma matrices, namely matrices \( i \gamma_i^{(g)} \) and \( \gamma_j^{(g)} \)  for $1 \leqslant i \leqslant p$ and $p+1 \leqslant j \leqslant p+q$, respectively, with $g={[(p+q)/2]}$.
For example, the Clifford algebra \( Cl_{3, 1} \) is represented by the 4×4 matrice $i\gamma_{1,2,3}^{(2)}$ and $\gamma_4^{(2)}$.

\section{Tenfold classification of TIs and TSCs}
\label{appendix e}
In this appendix, we outline the standard tenfold classification, which categorizes TIs and TSCs based on their symmetry properties and spatial dimensions. 
According to the presence or absence of time-reversal ($T$), particle-hole $(P)$, and chiral ($C$) symmetries, all the gapped systems can be classified into tenfold symmetry classes, as shown in Table \ref{tab:classification}. The classification of TIs or TSCs for a given symmetry class and spatial dimension can be derived through the Dirac Hamiltonian method \cite{Kitaev2009,Morimoto2013,Chiu2013,Chiu2014,Chiu2014a}.

For a system in $d$ dimension, the general form of the Dirac Hamiltonian is given by:
\beqn
&& H(\bm k)=\sum_{j=0}^D m_j \bar{\gamma}_{j}+\sum_{i=1}^{d} k_i \gamma_i, \nonumber\\
&&\left\{\gamma_i, \gamma_j\right\}=2 \delta_{i j} \mathbb{I},\left\{\bar{\gamma}_i, \bar{\gamma}_j\right\}=2 \delta_{i j} \mathbb{I},\left\{\bar{\gamma}_i, \gamma_j\right\}=0,\nonumber\\
\eeqn
where the momentum vector is $\bm k=(k_1,\cdots,k_d)$  and $m_j$ denotes the Dirac mass.
 The topological classification of gapped systems for each symmetry class is completely determined by the existence of SPEMT for $H$. For a given symmetry class, if the matrix dimension of $H$ for $D=0$ is the same as that for $D=1$, then the SPEMT exists. Otherwise, SPEMT does not exist and nontrivial bulk topology emerges. Generally, if the  matrix dimension of $H$ for $D=n$ is different from that for $D=n+1$, then the SPEMT does not exist for Dirac Hamiltonian at $(d-n)$D boundary and nontrivial boundary topology emerges, as described
 in in Sec~\ref{III}. This unified criterion leads to full classification of bulk/boundary TIs and TSCs.

For symmetry classes A, the construction of Dirac Hamiltonian requires $(d+D+1)$  anti-commuting Gamma matrices. While for symmetry classes AIII, $(d+D+2)$  anti-commuting Gamma matrices are required to construct the Hamiltonian and the chiral symmetry operator.
 These matrices generate the complex Clifford algebras $Cl_{D+d+1}$ and $Cl_{D+d+2}$, respectively, for A and AIII symmetry classes. The minimum matrix dimension for $Cl_q$ is $n_q=2^{[q/2]}$. For class A, when $d$ is even, the minimum matrix dimensions for $D=0$ and $D=1$ differ, signaling the absence of SPEMT and therefore, a nontrivial first-order topological classification. A similar analysis generates the complete first-order classification for the  AIII symmetry class, as shown in the third rows of Table \ref{tab:classification}.

For the other eight real symmetry classes, $H$ respects either $P$ or $T$ symmetry. To construct the Hamiltonian and corresponding symmetries, we need to have the following elements: 
\beqn
\mathcal{S}=\{i,\ \tilde{\gamma}_0,\ \tilde{\gamma}_1,\cdots,\ \tilde{\gamma}_D,\ \gamma_1,\ \gamma_2,\cdots,\ \gamma_d,\ T,\ \text{and(or)}\ P\} \label{elements},\nonumber\\
\eeqn
where $i \in \mathcal{S}$ ensures a complex structure ($i^2 = -1$), and the inclusion of $T$ and/or $P$ depends on the specific symmetry class. Because of anti-linearity, we have
\beqn
&&\{i,T\}=0,\{i,P\}=0,[\bar{\gamma}_j,T]=0,\nonumber\\
&&\{\tilde{\gamma}_j,T\}=0, \{\gamma_i,T\}=0,
[{\gamma}_j,P]=0.
\eeqn
Under these constraints, the algebra $G_{\mathcal{R}}$ generated by $\mathcal{S}$ for a symmetry class $\mathcal{R}$ is isomorphic to a specific  real Clifford algebra as \cite{Kitaev2009,Morimoto2013,Chiu2013,Chiu2014,Chiu2014a}:
\begin{align}
G_{\text{D}}\cong &Cl_{2+D,1+d}, & G_{\text{DIII}}\cong &  Cl_{3+D,1+d}, \nonumber \\
G_{\text{AII}} \cong &Cl_{3+D,d},\ & G_{\text{CII}}\cong &Cl_{4+D,d}, \nonumber \\
G_{\text{C}}\cong & Cl_{2+d,1+D},\ & G_{\text{CI}}\cong & Cl_{2+d,2+D},  \nonumber \\ 
G_{\text{AI}}\cong & Cl_{1+d,2+D}, & G_{\text{BDI}}\cong &  Cl_{1+d,3+D} , \label{iso}
\end{align}
By using these isomorphic relations and the representation theory of real Clifford algebras, the full classification of bulk TIs and TSCs for the real symmetry classes has been derived in Refs. \onlinecite{Kitaev2009,Morimoto2013,Chiu2013,Chiu2014,Chiu2014a}, as summaried in Table~\ref{tab:classification}. This framework also determines the minimal matrix dimension of the Dirac Hamiltonian for describing TIs and TSCs. By exploiting the  matrix dimensional correspondence between bulk and boundary Dirac Hamiltonians, we can further obtain lattice models of boundary TIs and TSCs of the full AZ symmetry classes.

% According to these isomorphic relationsand representation theory of real Clifford algebra, we can obtain full classification of TIs and TSCs for the real symmetry class, as shown in Table \ref{tab:classification}. Through this classification, we can also obtain the matrix dimension of the Dirac Hamiltonaian for describing TIs and TSCs in each symmetry and dimension. Building on the dimension correspondence between bulk and boundary  Dirac Hamiltoninas, we can generally obtain boundary TIs and TSCs described by lattice models. }

\begin{table}[tb]
\centering 
\begin{ruledtabular}
\begin{tabular}{c|ccc|cccccccc} 
   $\mathrm{class} \backslash d $ & $T$ & $P$ & $C$ & 0 & 1 & 2 & 3 & 4 & 5 & 6 & 7   \\  \hline
  A & 0 & 0 &0 & $\mathbb{Z}$ & 0 & $\mathbb{Z}$ & 0 & $\mathbb{Z}$ & 0 & $\mathbb{Z}$ & 0               \\
  AIII & 0 & 0 & 1 & 0 & $\mathbb{Z}$ & 0 & $\mathbb{Z}$ & 0 & $\mathbb{Z}$ & 0 & $\mathbb{Z}$             \\  
  \hline 
  AI & $+$ & 0 & 0  & $\mathbb{Z}$ & 0 & 0 & 0 & $2\mathbb{Z}$ & 0 & $\mathbb{Z}_2$ & $\mathbb{Z}_2$       \\
  BDI    & $+$ & $+$ & 1   & $\mathbb{Z}_2$ & $\mathbb{Z}$ & 0 & 0 & 0 & $2\mathbb{Z}$ & 0 & $\mathbb{Z}_2$   \\
  D & 0 & $+$ & 0  & $\mathbb{Z}_2$ & $\mathbb{Z}_2$ & $\mathbb{Z}$ & 0 & 0 & 0 & $2\mathbb{Z}$ & 0        \\
  DIII & $-$ & $+$ & 1 & 0 & $\mathbb{Z}_2$ & $\mathbb{Z}_2$ & $\mathbb{Z}$ & 0 & 0 & 0 & $2\mathbb{Z}$      \\
  AII & $-$ & 0 & 0 & $2\mathbb{Z}$ & 0 & $\mathbb{Z}_2$ & $\mathbb{Z}_2$ & $\mathbb{Z}$ & 0 & 0 & 0       \\
  CII & $-$ & $-$ & 1  & 0 & $2\mathbb{Z}$ & 0 & $\mathbb{Z}_2$ & $\mathbb{Z}_2$ & $\mathbb{Z}$ & 0 & 0      \\
  C & 0 & $-$ & 0 & 0 & 0 & $2\mathbb{Z}$ & 0 & $\mathbb{Z}_2$ & $\mathbb{Z}_2$ & $\mathbb{Z}$ & 0         \\
  CI & $+$ & $-$ & 1& 0 & 0 & 0 & $2\mathbb{Z}$ & 0 & $\mathbb{Z}_2$ & $\mathbb{Z}_2$ & $\mathbb{Z}$    
  \\   
\end{tabular}
\end{ruledtabular}
\caption{
Periodic table of first-order TIs and TSCs.
  This table is adopted from Ref.~\onlinecite{Chiu2016}.
\label{tab:classification}
}
\end{table}

\section{Analytical solution of boundary states}
\label{Appendix B}
Here we derive the zero-energy states solution in Eqs.~\eqref{dw1}, \eqref{wf1}, and \eqref{wf2}. We also derive the analytical solution of gapless boundary states of the Hamiltonian ${H}_{d}^{(n+1)}$ in Eq.~\eqref{ho}.

\subsection{Analytical solution of zero-energy states}
We rewrite the 1D domain wall Hamiltonian in Eq.~\eqref{hd0} as
\beqn
{H}_1^{(1)}(r_d)=m_1(r_d)\gamma_0^{(g)} -i\gamma_d^{(g)}\partial_{r_d},
\eeqn
where $m_1(r_d)=m_1>0$ in the region $r_d > 0$ and $m_1(r_d)=-m_1<0$ in the region $r_d <0$. The zero-energy solution of ${H}_1(r_d)$ satisfies the equation
\beqn
\gamma_0(m_1(r_d) -i\gamma_0^{(g)}\gamma_d^{(g)}\partial_{r_d})\Psi(r_d)=0,
\label{b1}
\eeqn
which indicates that $\Psi(r_d)$ is the eigenstate of $i\gamma_0^{(g)}\gamma_d^{(g)}$. Therefore, we set $\Psi(r_d)$ with the form
\beqn
\Psi(r_d)=\mathcal{N}f(r_d)\psi_{\alpha},
\eeqn
where $\mathcal{N}$ is the normalization factor, $f(r_d)$ is the real space wave function, and $\psi_{\alpha}$ is a spinor.
Given the boundary condition $\Psi(r_d=\pm \infty)=0$, the solution 
of Eq.~\eqref{b1} is
\beqn
\Psi_{-}(r_d)=\mathcal{N}e^{-\int_{0}^{r_d}m_1(r_d^{\prime})dr_d^{\prime}}\psi_{-},
\label{b2}
\eeqn
where $i\gamma_0^{(g)}\gamma_d^{(g)}\psi_{-}=-\psi_{-}$. Because $i\gamma_0\gamma_d$ has $2^{g-1}$ eigenstates with eigenvalue $-1$, ${H}_1(r_d)$ has $2^{g-1}$ zero-energy states localized close to $r_d=0$.

We rewrite the $n$D domain wall Hamiltonian presented in Eq.~\eqref{hd} as 
\beqn
{H}_n^{(n)}({\bm r_{n}})=\sum_{j=q+1}^{d}-i\partial_{r_j}\gamma_j^{(g)}+m_{j-q}\gamma_{j+n+1}^{(g)},
\label{b3}
\eeqn
where $m_j(r_j)=m_j>0$ for $r_j>0$ and $m_j(r_j)=-m_j<0$ for $r_j<0$, with $j=q+1,\cdots,d$, and $\bm r_n=(r_{q+1},\cdots,r_d)$ with $q=d-n$.
Given that ${H}_{n}^{(n)}({\bm r_n})$ can be separated into $r_{q+1},\cdots, r_{d}$ independent parts, we set the ansatz for the zero-energy solution as
\beqn
\Psi_{\alpha}(\bm r_n)=\mathcal{N}\prod_{j=q+1}^{d}f(r_j)\psi_{\alpha}.
\eeqn
The equation ${H}_n^{(n)}({\bm r_n})\Psi_{\alpha}(\bm r_n)=0$ can be satisfied when
\beqn
&&h_j(r_j)\Psi_{\alpha}(\bm r_n)=0,\nonumber\\
&&h_j(r_j)=-i\partial_{r_j}\gamma_j^{(g)}+m_{j-q}\gamma_{j+n+1}^{(g)},
\eeqn
where $j=q+1,\cdots,d$. According to the 1D domain wall zero-energy solution presented in Eq.~\eqref{b2}, we know that $\Psi(\bm r_n)$ is the eigenstate of $i\gamma_j^{(g)}\gamma_{j+n+1}^{(g)}$ with eigenvalue $-1$. Thus, $\Psi(\bm r_n)$ should be the common eigenstates of matrices $\{\mathcal{C}_{m+1},\cdots,C_{j},\cdots,\mathcal{C}_d\}$, where $\mathcal{C}_j=i\gamma_j^{(g)}\gamma_{j+n+1}^{(g)}$ and
\beqn
\mathcal{C}_j\Psi(\bm r_n)=-\Psi(\bm r_n).
\eeqn
The equation $h_s(r_s)\Psi(\bm r_n)=0$ further requires that
\beqn
f(r_j)=e^{-\int_{0}^{r_s}m_s(r_s^{\prime})dr_s^{\prime}}.
\eeqn
Therefore, ${H}_n^{(n)}({\bm r_n})$ hosts the zero-energy states 
\beqn
\Psi_{-,\cdots,-}(\bm r_n)=\mathcal{N}_1\prod_{j=q+1}^{d}e^{-\int_{0}^{r_s}m(r_s^{\prime})dr_s^{\prime}}\psi_{-,\cdots,-},
\eeqn
where spinor $\psi_{-,\cdots,-}$ satisfies $\mathcal{C}_j\psi_{-,\cdots,-}=-\psi_{-,\cdots,-}$. Under this restriction, matrices $\{\mathcal{C}_{q+1},\cdots,\mathcal{C}_d\}$ has $2^{(g-n)}$ eigenstates with eigenvalue $-1$. Therefore, ${H}_n^{(n)}({\bm r_n})$ has $2^{(g-n)}$ corner states localized close to $\bm r_n=0$.

We rewrite the Hamiltonian in Eq.~\eqref{BT} as
\beqn
&&{H}_{n}^{(n)}(\bm k_{n})=\sum_{j=q+1}^{d}h_{j}(k_j),\nonumber\\
&&h_{j}(k_j)=k_j\gamma_{j}^{(g)}
+M_{j-q}(k_{j})\gamma_{j+n+1}^{(g)},
\eeqn
where $M_{j-q}(k_j)=t_j+\cos k_j$ with $|t_j|<|1|$.
${H}_{n}^{(n)}$ can be regarded as the $n$D generalization of the BBH model, whose topological properties were studied in Ref.~\onlinecite{Luo2023a}. $h_j(k_j)$ is the 1D extended SSH model, of which the zero-energy states can be derived by a low-energy expansion (see Ref.~\onlinecite{Luo2023a} )
\beqn
X_{z_j}^{(j)}(r_j)=f_{z_j}^{(j)}(r_j)\psi_{z_j}^{(j)},
\label{IIIB1}
\eeqn
where spinor $\psi_{z_j}^{(j)}$ satisfies $\mathcal{C}_j\psi_{z_j}^{(j)}=z_j\psi_{z_j}^{(j)}$, with the eigenvalue $z_j=\pm 1$. The zero-energy states $X_{-}^{(j)}(r_j)$ and $X_{+}^{(j)}(r_j)$ are localized close to ends $r_j=0$ and $r_j=L$ , respectively, with $L$ the length along the $r_j$ direction. Real-space wave function $f_{z_j}^{j}(r_j)$ is found to be
\beqn
&&f_{-}^{j}(r_j)=\mathcal{N}(e^{\xi_{-}^{1}r_j}-e^{\xi_{-}^{2}r_j}),\nonumber\\
&&f_{+}^{j}(r_j)=\mathcal{N}(e^{\xi_{+}^1(r_j-L)}-e^{\xi_{+}^2(r_j-L)}).
\eeqn
where $ \xi_{z_j}^{1,2}={z_j \pm \sqrt{1-2(t_j+1)}}$.

Through a similar analysis, it can be demonstrated that ${H}_{n}^{(n)}(\bm r_{n})$ has $2^{g}$ zero-energy corner states and can be expressed as
\beqn
\Psi_{z_{q+1},\cdots,z_d}(\bm r_n)=\prod_{j=q+1}^df_{z_j}^{(j)}(r_j)\psi_{z_{q+1},\cdots,z_d},
\label{cwf}
\eeqn
where spinor $\psi_{z_{q+1},\cdots,z_d}$ is the common eigenstates of matrices $\{\mathcal{C}_{q+1},\cdots,\mathcal{C}_j,\cdots,\mathcal{C}_d\}$ and is defined by
\beqn
\mathcal{C}_j\psi_{z_{q+1},\cdots,z_d}=z_s\psi_{z_{q+1},\cdots,z_d}.
\eeqn

\subsection{Analytical solution of gapless boundary states of  ${H}_{d}^{(n+1)}$ in Eq.~\eqref{ho}.}
We rewrite the Hamiltonian in Eq.~\eqref{ho} as
\beqn
{H}_d^{(n+1)}(\bm k_d)&=\sum_{i=1}^{d}\sin k_i\gamma_i+M_0(\bm k_q)\gamma_{d+1}\nonumber\\
&+\sum_{j=1}^{n}M_j( k_{q+j})\gamma_{j+d+1},
\eeqn
where $M_0(\bm k_q)=\mathcal{M}+\sum_{i=1}^{q}(1-\cos k_i)$ with $-2<\mathcal{M}<0$ and $q=d-n$, and $M_{j}=(t_j+\cos k_{q+j})$. We first expand $M_0(\bm k_q)$ at $\bm k_q=0$, which leads to $M_0(\bm k_q)=\mathcal{M}+\sum_{i=1}^{q}k_i^2/2$.
We further take the open boundary condition along the $r_{q},\cdots,r_{d}$ directions.
In this case, ${H}_d^{(n+1)}$ can be separated into two parts as
\beqn
\label{BD1}
&&{H}_d^{(n+1)}(\bm k_{q-1},\bm r_{n+1})={H}_{q-1}^{(0)}(\bm k_{q-1})+{H}_{n+1}^{(n+1)}(\bm r_{n+1}),\nonumber\\
&&{H}_{q-1}^{(0)}(\bm k_{q-1})=\sum\limits_{i=1}^{q-1}k_i\gamma_{i}^{(g)},\nonumber\\
&&{H}_{n+1}^{(n+1)}(\bm r_{n+1})=\sum_{j=q}^{d}h_j(r_j),\nonumber\\
&&h_j(r_j)=-i\partial_{r_j}\gamma_{j}^{(g)}
+M_{j-q}(-i\partial_{r_j})\gamma_{j+n+1}^{(g)},
\label{b4}
\eeqn
where 
$\bm k_{q-1}=(k_1,\cdots,k_{q-1})$ and $\bm r_{n+1}=(r_q,\cdots,r_{d})$, and the insignificant terms involving $k_{1,\cdots,q-1}$ in $M_0$ has been omitted. ${H}_{n+1}^{(n+1)}$ can be viewed as the $(n+1)$D generalization of the BBH model and hosts $2^{(g-n-1)}$ corner states at each corner in the hyper cubic space. These corner states are the common eigenstate of matrices $\{\mathcal{C}_q,\cdots,\mathcal{C}_j,\cdots,\mathcal{C}_d\}$ and take a similar wave function form as that in Eq.~\eqref{cwf}, with $\mathcal{C}_j=i\gamma_j^{(g)}\gamma_{j+n+1}^{(g)}$.

Since ${H}_d^{(n+1)}$ can be separated into $\bm k_{q-1}$ and $\bm r_n$ independent parts under the low-energy expansion, the gapless boundary states of ${H}_d^{(n+1)}$ are the common eigenstates of ${H}_{n+1}^{(n+1)}$ and ${H}_{q-1}^{(0)}(\bm k_{q-1})$, 
\beqn
&&\Phi_{z_{q}\cdots z_d}(\bm k_{q-1},\bm r_{n+1})=\prod_{j=q}^{d}f_{z_j}^{j}(r_j)P_{z_{q}\cdots z_d}\phi(\bm k_{q-1}),\nonumber\\
&&{H}_{q-1}^{(0)}(\bm k_{q-1}){\phi}(\bm k_{q-1})=E(\bm k_{q-1}){\phi}(\bm k_{q-1}),\nonumber\\
&&P_{z_{q}\cdots z_d}=\prod_{j=q}^{d}P_{z_j},P_{z_j}=(1+z_j\mathcal{C}_j)/2,
\label{wf}
\eeqn
where $z_j$ is the eigenvalue of $\mathcal{C}_j$ with $z_j=\pm 1$, and $E(\bm k_{q-1})=\pm\sqrt{\sum_{i=1}^{q-1}k_i^2}$ is gapless at $\bm k_{q-1}=0$ .

\section{Proof of the one-to-one correspondence between the SPEMT for bulk and boundary Hamiltonians}
\label{Appendix C}
We show that the bulk and boundary Hamiltonians belong to the same symmetry class. Furthermore, the SPEMT for bulk and boundary Hamiltonians has a one-to-one correspondence.

We rewrite the Hamiltonian in Eq.~\eqref{2m} as
\beqn
{H}_d^{(2)}(\bm k_d)=\sum_{i=1}^d k_d \gamma_d^{(g)}+m_0 \gamma_{d+1}^{(g)}+m_1\gamma_{d+2}^{(g)}.
\label{c1}
\eeqn
We consider the Gamma matrices representation
\beqn
\gamma_d^{(g)}=\sigma_2\otimes I^{(g-1)}, \gamma_{d+1}^{(g)}=\sigma_1\otimes I^{(g-1)}.
\eeqn
Then ${H}_d^{(2)}(\bm k_d)$ can be rewritten as
\beqn
{H}_d^{(2)}(\bm k_d)=\sigma_3\otimes \tilde{H}_{d-1}^{(1)}(\bm k_{d-1})+k_d \gamma_d^{(g)}+m_0\gamma_{d+1}^{(g)},
\label{c3}
\eeqn
where
\beqn
\tilde{H}_{d-1}^{(1)}(\bm k_{d-1})=\sum_{i=1}^{d-1} k_d\tilde{ \gamma}_d^{(g-1)}+m_1\tilde{\gamma}_{d+2}^{(g-1)},
\eeqn
By following Eqs.~\eqref{hd0}-\eqref{bh}, one can show that $\tilde{H}_{d-1}^{(1)}(\bm k_{d-1})$ is the exact boundary Hamiltonian when taking the open boundary condition along the $k_d$ direction for ${H}_d^{(2)}$. Therefore, Eq.~\eqref{c3} can be understood as a Hamiltonian map, which bridges the bulk Hamiltonian ${H}_d^{(2)}$ and boundary Hamiltonian $ \tilde{H}_{d-1}^{(1)}$ by a dimension reduction. Supposing that $ \tilde{H}_{d-1}^{(1)}$ respects  $\tilde{\mathcal{T}}$,  $\tilde{\mathcal{P}}$, and  $\tilde{\mathcal{C}}$ symmetries, one can show that ${H}_d^{(2)}$ respects the local symmetries
\beqn
\mathcal{T}=\sigma_0\otimes\tilde{\mathcal{T}}, \mathcal{C}=\sigma_z\otimes \tilde{\mathcal{C}},\mathcal{P}_{d}=\sigma_z\otimes \tilde{\mathcal{P}},
\eeqn
where
\beqn
&&[\gamma_{d+1,d+2}^{(g)},\mathcal{T}]=0,\quad\{\gamma_{1,\cdots,d}^{(g)},\mathcal{T}\}=0,\nonumber\\
&&\{\gamma_{d+1,d+2}^{(g)},\mathcal{P}\}=0,\quad[\gamma_{1,\cdots,d}^{(g)},\mathcal{P}]=0,\nonumber\\
&&\{\gamma_{d+1,d+2}^{(g)},\mathcal{C}\}=0,\quad\{\gamma_{1,\cdots,d}^{(g)},\mathcal{C}\}=0.\nonumber\\
&&[\tilde{\gamma}_{d+2}^{(g-1)},\tilde{\mathcal{T}}]=0,\quad\{\tilde{\gamma}_{1,\cdots,d-1}^{(g-1)},\tilde{\mathcal{T}}\}=0,\nonumber\\
&&\{\tilde{\gamma}_{d+2}^{(g-1)},\tilde{\mathcal{P}}\}=0,\quad[\tilde{\gamma}_{1,\cdots,d-1}^{(g-1)},\tilde{\mathcal{P}}]=0,\nonumber\\
&&\{\tilde{\gamma}_{d+2}^{(g-1)},\tilde{\mathcal{C}}\}=0,\quad\{\tilde{\gamma}_{1,\cdots,d-1}^{(g-1)},\tilde{\mathcal{C}}\}=0.
\eeqn
By contrast, the presence of  $\mathcal{T}$, $\mathcal{P}$, or $\mathcal{C}$ symmetries of ${H}_d^{(2)}$ indicate that $ \tilde{H}_{d-1}^{(1)}$ respects the $\tilde{\mathcal{T}}$, $\tilde{\mathcal{P}}$, or $\tilde{\mathcal{C}}$ symmetries. Therefore, there is a one-to-one correspondence between the
local symmetries of $ \tilde{H}_{d-1}^{(1)}$ and ${H}_d^{(2)}$, which implies that $ \tilde{H}_{d-1}^{(1)}$ and ${H}_d^{(2)}$ belong to an identical symmetry class.

Based on the Hamiltonian map given by Eq.~\eqref{c3}, we can show that the SPEMTs of $ \tilde{H}_{d-1}^{(1)}$ and ${H}_d^{(2)}$ also allow a one-to-one correspondence. If $ \tilde{H}_{d-1}^{(1)}$ allows a SPEMT, labelled as $m_2\tilde{\gamma}_{d+3}^{(g-1)}$, then ${H}_d^{(2)}$ allows a SPEMT $m_2{\gamma}_{d+3}^{(g)}$, with
${\gamma}_{d+3}^{(g)}=\sigma_3\otimes\tilde{\gamma}_{d+3}^{(g-1)}$. Inversely, if ${H}_d^{(2)}$ allows a SPEMT $m_2{\gamma}_{d+3}^{(g)}$, then ${\gamma}_{d+3}^{(g)}$ must takes the form $\sigma_3\otimes\tilde{\gamma}_{d+3}^{(g-1)} $ because of $\{{\gamma}_{d+3}^{(g)},\gamma_{d,d+1}^{(g)}\}=0$, which leads to $\{\tilde{\gamma}_{d+3}^{(g-1)},\tilde{H}_{d-1}^{(1)}\}=0$. Therefore, there is a one-to-one correspondence for the SPEMT of $ \tilde{H}_{d-1}^{(1)}$ and ${H}_d^{(2)}$.

To obtain the Hamiltonian $H_{d}^{(n+1)}$ presented in Eq.~\eqref{gd}, we consider the iterative Hamiltonian map
\beqn
&&{H}_{q+j}^{(1+j)}=\sigma_3\otimes {H}_{q+j-1}^{(j)}+k_{q+j}\gamma_{q+j}^{(g-n+j)}+m_j \gamma_{d+j}^{(g-n+j)},\nonumber\\
&&\gamma_{q+j}^{(g-n+j)}=\sigma_2\otimes I^{(g-n+j)},\gamma_{d+j}^{(g-n+j)}=\sigma_1\otimes I^{(g-n+j)},\nonumber\\
\label{hm}
\eeqn
where $j\geq 1$. Similarly, Hamiltonians ${H}_{q+j}^{(1+j)}$ and ${H}_{q+j-1}^{(j)}$ belong to the same symmetry class and the SPEMT of them have a one-to-one correspondence. Moreover, when taking the open boundary condition along the $k_{q+j}$ direction for ${H}_{q+j}^{(1+j)}$, we obtain the boundary Hamiltonian ${H}_{q+j-1}^{(j)}$. Inputting the initial Hamiltonian $H_{q}^{(1)}$, 
 ${H}_{d}^{(n+1)}$ can be produced by performing the Hamiltonian map given by Eq.~\eqref{hm}  $n$ times, corresponding to $j=1,\cdots,n$. Therefore, by taking the open boundary conditions along the $k_{q+1},\cdots,k_d$ directions,  ${H}_{d}^{(n+1)}$ can be connected to the boundary Hamiltonian ${H}_{q}^{(1)}$ and the SPEMT of them
has a one-to-one correspondence.

%This framework relies on three fundamental symmetries: \textbf{time-reversal symmetry (TRS)}, \textbf{particle-hole symmetry (PHS)}, and \textbf{chiral symmetry (CS)}. These symmetries define ten distinct symmetry classes, each associated with a specific Clifford algebra. The classification ensures that no additional symmetries permit mass terms that could destabilize the topological phases. Additionally, we provide the minimal matrix dimensions of the Dirac Hamiltonian for each class in spatial dimensions \( d = 1, 2, 3 \).

\section{The derivation of Eq.~\eqref{bdh1} }
\label{Appendix D}
We derive the surface Hamiltonian presented in Eq.~\eqref{bdh1}. We perform the surface projection analysis by decomposing $H_{7}(\bm k)$ (Eq.~\eqref{TI}) into two parts
\beqn
&&H_{7}(\bm k)={H}_2^{(1)}(\bm k_{||})+{H}_{1}^{(1)}( k_{3}),\nonumber\\
&&{H}_2^{(1)}(\bm k_{||})=\sin k_1\gamma_1^{(3)}+\lambda\sin k_2\gamma_2^{(3)}+\Delta(\bm k_{||})\gamma_4^{(3)},\nonumber\\
&&{H}_{1}^{(1)}( k_{3})=\sin k_3\gamma_3^{(3)}+(t+\cos k_3)\gamma_5^{(3)},
\eeqn
where the anti-commuting $8 \times 8$ Gamma matrices are
\beqn
&&\gamma_1^{(3)}=\tau_3\sigma_3s_1,\gamma_2^{(3)}=\tau_3\sigma_3s_2,\nonumber\\
&&\gamma_3^{(3)}=\tau_3\sigma_3s_3, \gamma_4^{(3)}=\tau_x\sigma_0s_0, \gamma_5^{(3)}=\tau_3\sigma_3s_0.
\eeqn
Taking the open boundary condition along the $k_3$ direction, ${H}_1^{(1)}(r_3)$ has eight zero-energy states, which are the eigenstates of the operator $\mathcal{C}=i\gamma_3^{(3)}\gamma_5^{(3)}$.  Note that $[\mathcal{C},{H}_2^{(1)}]=0$, the surface states of $H_{7}$ are the common eigenstates of $\mathcal{C}$ and ${H}_2^{(1)}$, which can be written as
\beqn
&&\Phi(\bm k_{||},r_3)=f(r_3)P_{+}\phi(\bm k_{||}),\nonumber\\
&&{H}_{2}^{(1)}(\bm k_{||})\phi(\bm k_{||})=E(\bm k_{||})\phi(\bm k_{||}).
\label{ss}
\eeqn
where the surface projection operator is $P_{+}=(1+ \mathcal{C})/2$ and $f(r_3)$ is the real space wave function of the zero-energy states of ${H}_1^{(1)}({r_3})$ localized close to $r_3=L$. The surface projection operator $P_{+}$ can be simplified as
\beqn
&&U^{\dagger}P_{+}U=(1-\tau_z)/2, \nonumber\\
&&U=\frac{\tau_x(\sigma_z+\sigma_0)s_0+\tau_0(\sigma_0-\sigma_z)s_0}{2}e^{-i\frac{\pi}{4}\tau_z\sigma_xs_z}.\nonumber\\
\eeqn
Projecting ${H}_2^{(1)}(\bm k_{||})$ onto the subspace defined by $P_{+}$ and picking up the non-zero block part, the surface Hamiltonian can be written as
\beqn
\tilde{{H}}_{2}^{(1)}&=&p_1^{T}U^{\dagger}{H}_{2}Up_1\nonumber\\
&=&-\Delta(\bm k_{||})\tilde{\tau}_y\tilde{s}_z-\lambda(\sin k_x\tilde{s}_y-\sin k_y\tilde{s}_x),
\eeqn  
where the $8\times 4$ matrix $p_1$ is defined as $[0,\sigma_0s_0]^{T}$ to extract the non-zero block of projector $P_{+}$, Pauli matrices $\tilde{\tau},\tilde{s}$ act on the subspace corresponding to the non-zero block of $P_{+}$. The time-reversal and particle-hole symmetries of $\tilde{{H}}_{2}^{(1)}$ are given by
\beqn
\tilde{\mathcal{T}}=&p_1^{T}U^{\dagger}is_yKUp_1=i\tilde{s}_yK,\nonumber\\
\tilde{\mathcal{P}}=&p_1^{T}U^{\dagger}\tau_ys_yKUp_1=-i\tilde{\tau}_x\tilde{s}_xK,
\eeqn
which satisfy $\tilde{\mathcal{T}}^2=-1$ and $\tilde{{P}}^2=1$.

\bibliography{reference}

\end{document}